\PassOptionsToPackage{pdfpagelabels=false, pageanchor=false, hyperfootnotes=false, plainpages=false,hypertexnames=false}{hyperref} 
\RequirePackage{rotating}
\documentclass[fleqn,usenatbib]{mnras}
\usepackage{pdflscape}
\usepackage[T1]{fontenc}
\usepackage{ae,aecompl}
\usepackage{graphicx}
\usepackage{amsmath, bm}
\usepackage{amssymb}
\usepackage{txfonts}
\usepackage{mathptmx}

\usepackage{textcomp}
\usepackage{gensymb}
\usepackage{etoolbox}
\usepackage{xspace}
\usepackage{paralist}
\usepackage{blindtext}
\usepackage[english]{babel}
\usepackage[usenames,dvipsnames, table]{xcolor}
\usepackage{caption}

\usepackage{tabularx}  

\hypersetup{colorlinks,linkcolor={NavyBlue},citecolor={NavyBlue},urlcolor={NavyBlue}}

\renewcommand{\vec}[1]{\boldsymbol{#1}}
\newcommand{\pc}{\,\rm{pc}\xspace}
\newcommand{\kpc}{\,\rm{kpc}\xspace}
\newcommand{\kms}{\,\rm{km\,s^{-1}}\xspace}
\newcommand{\kmskpc}{\,\rm{km\,s^{-1}\,kpc^{-1}}\xspace}
\mathchardef\mhyphen="2D

\newcommand{\nmagic}{NMAGIC\xspace}
\newcommand{\brava}{BRAVA\xspace}
\newcommand{\argos}{ARGOS\xspace}
\newcommand{\twomass}{2MASS\xspace}
\newcommand{\ukidss}{UKIDSS\xspace}
\newcommand{\glimpse}{GLIMPSE\xspace}
\newcommand{\ogle}{OGLE\xspace}
\newcommand{\vvv}{VVV\xspace}
\newcommand{\apogee}{APOGEE\xspace}
\newcommand{\gibs}{GIBS\xspace}

\newcommand{\masyr}{\,\rm{mas\,yr^{-1}}\xspace}

\newcommand{\Msun}{\,\rm{M}_{\odot}\xspace}

\newcommand{\Gyr}{\,\rm{Gyr}}

\newcommand{\vel}{\textrm{v}}

\newcommand\Tstrut{\rule{0pt}{2.9ex}}         
\newcommand\Bstrut{\rule[-2.9ex]{0pt}{0pt}}

\title[Dynamical modelling of the Galactic bar]{Dynamical modelling of the Galactic bulge and bar: \\ the Milky Way's bar pattern speed, stellar, and dark matter mass distribution}

\author[M. Portail et al.]
  {Matthieu~Portail$^1$\thanks{E-mail:portail@mpe.mpg.de}, Ortwin~Gerhard$^1$\thanks{E-mail:gerhard@mpe.mpg.de}, Christopher~Wegg$^1$ and Melissa~Ness$^2$\\
  $^1$ Max-Planck-Institut f\"{u}r Extraterrestrische Physik, Gie\ss enbachstra\ss e, D-85741 Garching, Germany\\
  $^2$ Max-Planck-Institut  f\"{u}r  Astronomie,  K\"{o}nigstuhl  17, D-69117 Heidelberg, Germany}

\date{Accepted 2016 October 31. Received 2016 October 28; in original form 2016 August 28}

\pubyear{2016}
\begin{document}
\phantomsection\label{firstpage}
\pagerange{\pageref{firstpage}--\pageref{lastpage}}
\maketitle

\begin{abstract}
We construct a large set of dynamical models of the galactic bulge, bar and inner disk using the Made-to-Measure method. Our models are constrained to match the red clump giant density from a combination of the \vvv, \ukidss and \twomass infrared surveys together with stellar kinematics in the bulge from the \brava and \ogle surveys, and in the entire bar region from the \argos survey. We are able to recover the bar pattern speed and the stellar and dark matter mass distributions in the bar region, thus recovering the entire galactic effective potential. We find a bar pattern speed of $39.0 \pm 3.5 \,\rm{km\,s^{-1}\,kpc^{-1}}$, placing the bar corotation radius at $6.1 \pm 0.5 \rm{kpc}$ and making the Milky Way bar a typical fast rotator. We evaluate the stellar mass of the long bar and bulge structure to be $M_{\rm{bar/bulge}} = 1.88 \pm 0.12 \times 10^{10} \, \rm{M}_{\odot}$, larger than the mass of disk in the bar region, $M_{\rm{inner\ disk}} = 1.29\pm0.12 \times 10^{10} \, \rm{M}_{\odot}$.  The total dynamical mass in the bulge volume is $1.85\pm0.05\times 10^{10} \, \rm{M}_{\odot}$. Thanks to more extended kinematic data sets and recent measurement of the bulge IMF our models have a low dark matter fraction in the bulge of $17\%\pm2\%$. We find a dark matter density profile which flattens to a shallow cusp or core in the bulge region. Finally, we find dynamical evidence for an extra central mass of $\sim0.2\times10^{10} \,\rm{M}_{\odot}$, probably in a nuclear disk or disky pseudobulge.
\end{abstract}

\begin{keywords}
methods: numerical -- Galaxy: bulge -- Galaxy: kinematics and dynamics -- Galaxy: structure -- Galaxy: centre
\end{keywords}

\section{Introduction}
\label{section:introduction}

Although it is well established that the Milky Way hosts a central barred bulge which causes non-axisymmetric gas flow \citep{Peters1975, Binney1991} and asymmetries in the near-infrared light \citep{Blitz1991, Weiland1994} and star counts \citep{Nakada1991, Stanek1997}; our understanding of this structure has dramatically improved in the last decade. The discovery of the so-called split red clump in the galactic bulge \citep{Nataf2010, McWilliam2010} and the later 3D mapping of the bulge density by \citet{Wegg2013} showed that the galactic bulge has a boxy/peanut (B/P) shape, similarly to bulges formed in N-body models by buckling of a vertically unstable stellar bar \citep{Combes1990, Martinez-Valpuesta2006}.

The existence of the bar outside of the bulge initially revealed by \citet{Hammersley1994} has been subject to controversy as to whether it is a separate structure from the bulge. First studies indicated a misalignment between the bulge and the bar, leading to the hypothesis of a double bar system in the inner Milky Way (\citealt{Benjamin2005}; \citealt{Lopez-Corredoira2005}; \citealt{Cabrera-Lavers2008}; but see also \citealt{Martinez-Valpuesta2011}). Recently, \citet[hereafter W15]{Wegg2015} demonstrated by combining the \vvv, \ukidss, \glimpse and \twomass catalogues that the galactic bulge smoothly segues into the long bar. Both  components appear at a similar angle, showing that the galactic bulge and the long bar in the Milky Way are consistent with being a single structure that became vertically thick in its inner part, similarly to the buckled bars of N-body models.

In \citet[hereafter P15]{Portail2015a}, we constructed dynamical models of the galactic bulge by combining the 3D density of Red Clump Giants (RCGs) in the bulge from \citet{Wegg2013} with bulge kinematics from the \brava survey \citep{Rich2007, Kunder2012} using the Made-to-Measure method. In this paper, we extend the Made-to-Measure modelling to the entire bar region by taking advantage of the recent measurement of \hyperlink{W15}{W15} on the bar outside the bulge together with stellar kinematics in the bar region from the \argos survey \citep{Freeman2013, Ness2013a}. The goal is to combine stellar density and kinematics under the constraint of dynamical equilibrium in order to recover the effective potential in the bar region, i.e. the stellar and dark matter mass distribution together with the bar pattern speed.

Extending the modelling from the bulge to the entire bar region is not a straightforward task. The galactic long bar extends about $5\kpc$ from the Galactic Centre (GC) as shown by \hyperlink{W15}{W15} and thus reaches radii where the outer disk contribution to the potential is important. Hence, modelling the bar region also requires modelling the disk potential into which the bar is embedded. In addition, the dark matter contribution to the radial force increases with galactocentric radius reaching about $50\%$ at the solar radius \citep{Read2014}. As a consequence, the global mass distribution of Galaxy has to be taken into account in order to produce a good model of the galactic bar region. The building of good initial conditions for the Made-to-Measure modelling is particularly challenging.

The paper is organized as follows. In \autoref{section:M2MMethod}, we briefly describe the Made-to-Measure method and the problem posed by the initial conditions. In \autoref{section:staticMWModels}, we construct a static density model of the inner $10\kpc$ for the Galaxy by combining the current knowledge of the bulge, bar, disk and dark matter density. This density model is used in \autoref{section:tailoringInitialConditions} to tailor a set of N-body models with different bar pattern speeds that already broadly match the Milky-Way mass distribution. In \autoref{section:nmagicConstraints}, we discuss the different data sets used in this study to constrain the models and summarize the modelling procedure in \autoref{section:M2Mfitting}. In \autoref{section:bulgeDynamics}, we show the effect of the main modelling parameters on the bulge dynamics, and in \autoref{section:Results} we analyse a large number of models, recover the bar pattern speed of the Milky Way and identify our best model. In \autoref{section:stellarAndDarkMatterMass}, we summarize our constraints on the stellar and dark matter mass distribution that arises from our models, and discuss our results in the light of other works in \autoref{section:discussion}. We finally conclude in \autoref{section:conclusion}. The impatient reader can read first \hyperref[section:Results]{Sections~\ref{section:Results}} and \hyperref[section:stellarAndDarkMatterMass]{\ref{section:stellarAndDarkMatterMass}} where we use our modelling to recover the effective potential in the Galaxy. Our best model is the first non-parametric model of the entire bar region and may be made available upon request to the authors.

\section{Made-to-Measure modelling of the Galaxy}
\label{section:M2MMethod}

\subsection{M2M modelling}
\label{section:M2Mtheory}
Stellar dynamical equilibria for galaxies can be studied with moment-based methods \citep{Binney1990, Cappellari2009}, classical distribution function-based methods \citep{Dejonghe1984, Qian1994}, actions-based methods \citep{Binney2010, Sanders2013}, orbit-based methods \citep{Schwarzschild1979, Thomas2009} or with the Made-to-Measure (M2M) method \citep{Syer1996, DeLorenzi2007}. In these M2M models, an initial self-gravitating N-body model is used to provide a discrete sample of a distribution function reasonably close to the system of interest. This N-body model is then slowly adapted by modifying the weights of the N-body particles such as to make the model reproduce a given set of constraints. The N-body weights can hence be seen simultaneously as mass elements (N-body point of view), or as weights for the orbits traced by the particles (Schwarzschild's method point of view). The method was extended to allow the fitting of observational data and implemented as the \nmagic code by \citet{DeLorenzi2007}.
The M2M method has been used in both extragalactic \citep[e.g.][]{DeLorenzi2008, DeLorenzi2009, Das2011, Morganti2013, Zhu2014} and Galactic context (e.g. \citealt{Bissantz2004}; \citealt{Long2013}; \citealt{Hunt2014}; \hyperlink{P15}{P15}). We heavily modified the \nmagic code for the purpose of modelling barred disk galaxies.

Formally, the M2M method works as follows. Any observable $y$ of a system can be written in terms of the distribution function $f(\vec{z})$ of the system by 
\begin{equation}
\label{equation:generalObservable}
 y = \int K(\vec{z}) f(\vec{z}) \,d^6\vec{z}
\end{equation}
where $K$ is the kernel corresponding to the observable and $\vec{z}$ the phase-space vector. In the M2M method, $f(\vec{z})$ is discretely sampled via a set of $N$ particles with particle weights $w_i(t)$. \autoref{equation:generalObservable} is then evaluated by
\begin{equation}
\label{equation:observable}
 y(t) = \sum_{i=1}^N K(\vec{z}_i(t)) w_i(t)
\end{equation}
where $\vec{z}_i(t)$ is the phase-space coordinate of particle $i$ at time $t$.

The M2M method consists of adjusting the particle weights $w_i$ in order to maximize a given profit function $F$. This is achieved by a simple gradient descent in which the particle weights are evolved with time according to
\begin{equation}
  \label{equation:FOC1}
  \frac{dw_i}{dt} = \varepsilon w_i \frac{\partial F}{\partial w_i}
\end{equation}
and $\varepsilon$ is a numerical factor that sets the typical time-scale of the weight evolution. The profit function $F$ usually consists of a chi-square term, which drives the model towards the data, and an entropy term to regularize the particle model. We describe the M2M formalism in more detail in \autoref{section:M2Mformalism}.

Note that the M2M method only weights particles and does not have the ability to create new N-body orbits. It is thus a very efficient modelling technique provided all the N-body orbits required to fit the data are already present in the initial model. Building good initial conditions, including its dark matter component, is a major issue when modelling the inner $10\kpc$ of the Milky Way.

\subsection{The problem of the initial conditions}
\label{section:InitialConditions}

To model the galactic bar, we need an initial N-body model of a barred stellar disk that provides a first-guess discrete sampling of the final model. The classical way to build such initial conditions is to evolve a near-equilibrium N-body stellar disk in a live dark matter halo. During the evolution the disk becomes unstable and forms a bar, that later forms a B/P bulge thought the buckling instability mechanism \citep{Combes1990, Martinez-Valpuesta2006}. This process has been used widely in order to build models that can then be compared to data \citep{Martinez-Valpuesta2006, Athanassoula2007, Shen2010} but suffers from three major limitations:
\begin{itemize}
 \item[(i)] The evolution process is non-linear and very sensitive to initial conditions. As noted by \citep{Sellwood2009} large changes in the evolved bar model can result from simply changing the seed of the random number generator used in building the initial conditions.
 \item[(ii)] The bar properties such as pattern speed, bar length and bar strength cannot be easily controlled a priori. 
 \item[(iii)] The bar tends to be 3-5 disk scale-lengths long as already noted by \citet{Debattista2000} and \citet{Athanassoula2002}. This is much larger than what would be required to model the Milky Way where this ratio lies within $1.9 \pm 0.4$ \citep{BlandHawthorn2016}.
\end{itemize}

\begin{figure}
  \centering
  \includegraphics{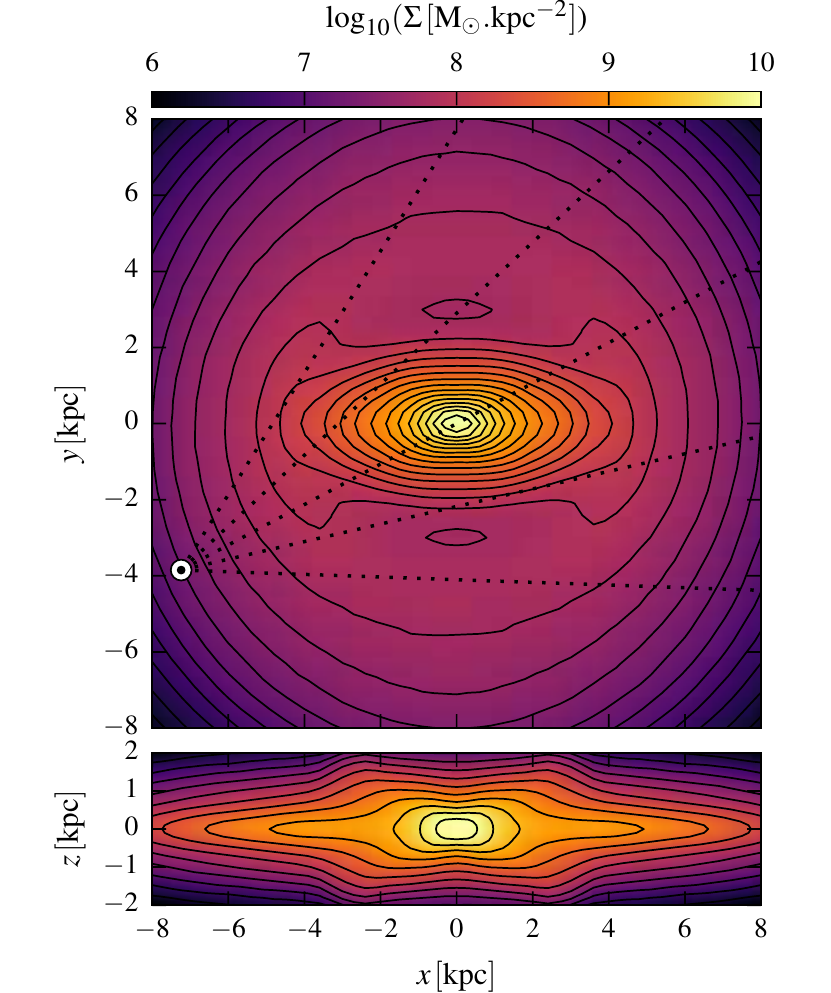}\\
  \caption{Face-on (upper) and side-on (lower) projections of the initial model M85 from \citet{Portail2015a}. The bar is $5\kpc$ long and at an angle of $\alpha=28\degree$ from the Sun-GC line of sight. The dotted lines originating from the Sun (dot symbol) indicate sight-lines with galactic longitudes $l=-30\degree, -15\degree, 0\degree, +15\degree$ and $+30\degree$.}
  \label{fig:initialModelM85}
\end{figure}

An example of such a buckled bar N-body model is the model M85 of \hyperlink{P15}{P15} shown in \autoref{fig:initialModelM85}. When scaled to a bar half-length of $5\kpc$, M85 has a very short disk scalelength of only $\sim1.2\kpc$, resulting in a very low contribution of the outer disk to the potential at the end of the bar region together with a lack of mass on long bar orbits. M85 was good enough to model only the bulge in \hyperlink{P15}{P15} but does not suit our purpose here to model the inner $10\kpc$ of the Galaxy. To get around the three limitations of pure N-body evolution shown above, we use a variant the M2M method to tailor initial conditions in a two-step process. In \autoref{section:staticMWModels}, we first create a mass density model of the Milky Way by adding together our best-guess densities for the bulge, bar, disk and dark halo. This density is then imprinted on M85 using a variant of the M2M method in \autoref{section:tailoringInitialConditions}, for different bar pattern speeds. At the end of this process, we obtain a family of N-body models with different bar pattern speeds that have broadly the right mass distribution and provide suitable initial conditions for modelling the inner $10\kpc$ of the Milky Way by fitting real data.

\section{Density model of the Galaxy for tailoring initial conditions}
\label{section:staticMWModels}
In this section, we construct a mass density model of the inner $10\kpc$ of the Galaxy by combining the 3D densities of the B/P bulge, bar, outer disk and dark matter halo. This density model is only used to tailor our initial conditions for the actual M2M modelling which is performed in \autoref{section:M2Mfitting}. Throughout the paper, we place the Sun in the galactic plane at a distance $R_0 = 8.2\kpc$ \citep{BlandHawthorn2016} from the GC. The bar is oriented at an angle of $\alpha=28\degree$ with the Sun-GC line of sight, consistent with the measurement of $27\degree \pm2\degree$ from the bulge RCGs \citep{Wegg2013} and the range $28\degree-33\degree$ measured by \hyperlink{W15}{W15} from the long bar RCGs. Following \citet{BlandHawthorn2016}, we assume that the local standard at rest (LSR) is on a circular orbit at $V(R_0) = 238\kms$, and a peculiar motion of the Sun in the LSR of $(U,V,W) = (11.1, 12.24, 7.25) \kms$ \citep{Schonrich2010}. This set of assumption predicts a solar tangential velocity of $250 \kms$, in good agreement with several recent measurements of $248 \pm 6 \kms$ \citep[from SEGUE data]{Schonrich2012}, $242^{+10}_{-3}\kms$ \citep[from APOGEE data]{Bovy2012}, $244\pm5\kms$ \citep[from RAVE data]{Sharma2014} and $251\pm5\kms$ \citep[from maser velocities]{Reid2014}.
All parameters of the density model are summarized in \autoref{table:ModelParameters}.

\subsection{Unified bulge and bar structure as traced by RCGs}
\label{section:bulgeAndBarModel}
Recently, \citet{Wegg2013} constructed the first non-parametric measurement of the 3D density of RCGs in the bulge. They took advantage of the narrow luminosity function of RCGs to directly deconvolve the extinction- and completeness-corrected magnitude distributions of bulge stars from the \vvv survey and obtain line of sight densities of RCGs. Combining the different lines of sight and assuming eight-fold symmetry, they produced a 3D density map of RCGs in a box of $(\pm 2.2 \times \pm 1.4 \times \pm 1.2)\kpc$ around the principal axes of the bulge. This map together with the \brava kinematics was later used in \hyperlink{P15}{P15} to construct a family of dynamical models of the galactic bulge. As bulge component, we adopt here the 3D density of the fitted model M85 of \hyperlink{P15}{P15} that reproduces the original RCG density very well, with the advantage of being smooth and complete in the plane where the direct measurement of the density was not possible because of extinction and crowding. 

Outside the bulge, \hyperlink{W15}{W15} combined the \vvv, \ukidss and \twomass surveys and showed that the bulge smoothly segues from its vertically extended B/P shape to the flat long bar. The bulge and long bar are shown in this later work to be consistent with forming a single structure, oriented at $\alpha = 28\degree$ from the Sun-GC line of sight. They estimated the long bar half-length to be $5\kpc$ and found evidence for an extra superthin bar component existing predominantly near the bar end. They finally fit a parametric model of the long bar density that once added to the fitted bulge model M85 of \hyperlink{P15}{P15} and convolved with the bulge luminosity function fits well the magnitude distribution of stars across the entire bulge and bar region. Consequently, we complement the bulge model described above using their best-fitting parametric densities of the thin long bar and superthin components. Note that due to their analysis method, the long bar density of \hyperlink{W15}{W15} does not include the inner disk, smooth background of stars filling the bar region. We add the inner disk density in \autoref{section:diskModel}.

Both the bulge and long bar density were measured using RCGs as tracers. Theoretical models by \citet{Salaris2002} show that for a $10\Gyr$ old stellar population, RCGs trace the stellar mass within 10\% for metallicities in the range $-1.5\leq [\rm{Fe/H}] \leq 0.2$. In the particular case of the galactic bulge and bar, \citet{Ness2013a} find from the \argos sample that $95\%$ of the stars enter this metallicity range. The age of the bulge is still under debate with pieces of contradictory evidence: photometric studies of the color-magnitude diagram \citep[CMD; ][]{Zoccali2003, Clarkson2008, Calamida2015} find that the bulge is older than $10\Gyr$ while spectroscopic age measurements of microlensed dwarfs find evidence for $4-5\Gyr$ old population among stars with $[\rm{Fe/H}]\geq-0.1$ \citep{Bensby2011, Bensby2013}. We assume here that the bulge and bar are $10\Gyr$ old, implying that the RCG density of the bulge and long bar considered above are proportional to the stellar density with expected variations of less than $10\%$. We call the proportionality factor between stellar mass density and RCG density \emph{mass-to-clump ratio} \footnote{Note that, as in \hyperlink{P15}{P15} our definition of the mass-to-clump also includes the red giant branch bump stars, the number of RCGs + red giant branch bump stars is better defined than the number of RCGs only (see \citealt{Wegg2013}).}, denoted as $\rm{M/n_{RCG}}$ by analogy with the mass to-light-ratio. We make the fiducial assumption that the bulge and the thin long bar have the same mass-to-clump ratio, as expected if they both formed at the same time. The origin of the superthin bar is still unclear as stated by \hyperlink{W15}{W15} but its stellar population is expected to be younger given its extremely short scaleheight ($45\pc$), and therefore it is likely to have a lower mass-to-clump ratio than the bulge. Assuming a constant star formation rate and a Kroupa initial mass function (IMF) as in \hyperlink{W15}{W15}, the mass-to-clump ratio of the superthin bar is a factor of 1.6 times smaller than that of a $10\Gyr$ old bulge.

\subsection{Empirical determination of the mass-to-clump ratio in the bulge}
\label{section:empiricalMassToClump}
The mass-to-clump ratio can be predicted by stellar population synthesis models using an IMF, a stellar age distribution and a metallicity distribution as in \hyperlink{P15}{P15}. The most recent measurement of the galactic bulge IMF is from \citet{Calamida2015} who used ultra-deep \emph{Hubble Space Telescope} (HST) photometry to recover the IMF in the mass range $0.15 \leq M/M_{\odot}\leq 1$. They find an IMF in good agreement with the Kroupa IMF \citep{Kroupa2001}, for which we computed in \hyperlink{P15}{P15} a mass-to-clump ratio of $\rm{M/n_{RCG}} = 984$ for a $10\Gyr$ old population. 

\begin{figure}
  \centering
  \includegraphics{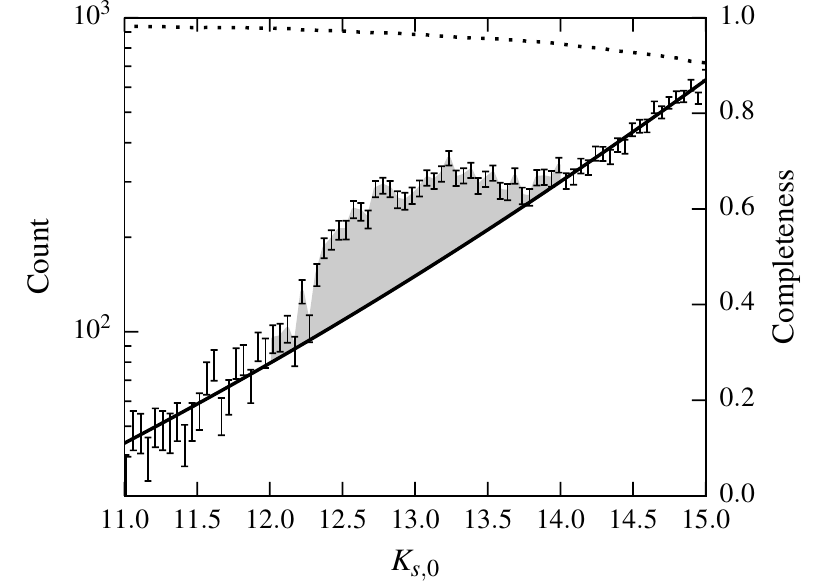}\\
  \caption{$K_s$-band extinction and completeness corrected magnitude distribution from the \vvv catalogue of \citet{Wegg2015} in a beam centred on the NICMOS field of radius $15\arcmin$. RCGs are identified as the excess above the background of stars. The dotted line indicates the completeness of the original \vvv catalogue as a function of magnitude.}
  \label{fig:MasstoClump}
\end{figure}

An alternative and more direct approach is to combine stellar mass measurements in some bulge field with the observed number of RCGs in that field, in analogy with the method of \citet{Valenti2015}. This approach is advantageous since it does not rely on stellar population models or parametrization of the IMF. \citet{Zoccali2000} used HST photometry in the NICMOS field and after cleaning for disk contamination they find a stellar mass in that field of $M_{\rm NICMOS} = 570\Msun$ (see the revision of the mass in the NICMOS field in \citealt{Valenti2015}). The NICMOS field has an area of only 408 square arcseconds and does not contain many giant stars. To improve the statistics on the number of RCGs, we follow \citet{Valenti2015} and consider a larger $15\arcmin$ beam centred on the NICMOS field and rescale the mass and number of RCGs by the ratio of the area of the two fields. We use the completeness- and extinction-corrected \vvv catalogue of \citet{Wegg2013} and identify RCGs statistically as the excess above the smooth background of stars in the extinction corrected magnitude distribution. \autoref{fig:MasstoClump} shows the magnitude distributions of \vvv stars in our larger field centred on the positions of the NICMOS fields with the identified RCGs above the background of stars.

With this approach, we find a mass-to-clump ratio of $\rm{M/n_{RCG}} = 1015$. We estimate the error on this figure of about 10\%, mostly due to systematic effect arising in defining the smooth background of stars on to which the RCGs sit (see \citealt{Wegg2013}). Note that the uncertainty on the low-mass end of the IMF due to unseen dwarfs has only a small effect on the mass-to-clump ratio. Variations from a sub-stellar slope of $-0.3$ \citep{Kroupa2001}, as also adopted in the revised mass of the NICMOS field \citep{Valenti2015}, to either a slope of $-1.33$ \citep{Zoccali2000} or a lognormal parametrization lead to variations of the mass-to-clump ratio of only $\pm3\%$. Our direct measurement of the mass-to-clump ratio is in good agreement with the predicted value of $984$ for a Kroupa or Calamida IMF. In all the following, we adopt the fiducial value of $\rm{M/n_{RCG}} = 1000$ for the main stellar population in the bulge and bar together with a lower value of $\rm{M/n_{RCG}} = 600$ for the superthin bar component. We show in \autoref{section:Results} the effect of a 10\% smaller or larger mass-to-clump ratio.

\subsection{Stellar disk}
\label{section:diskModel}

\begin{figure*}
  \centering
  \includegraphics{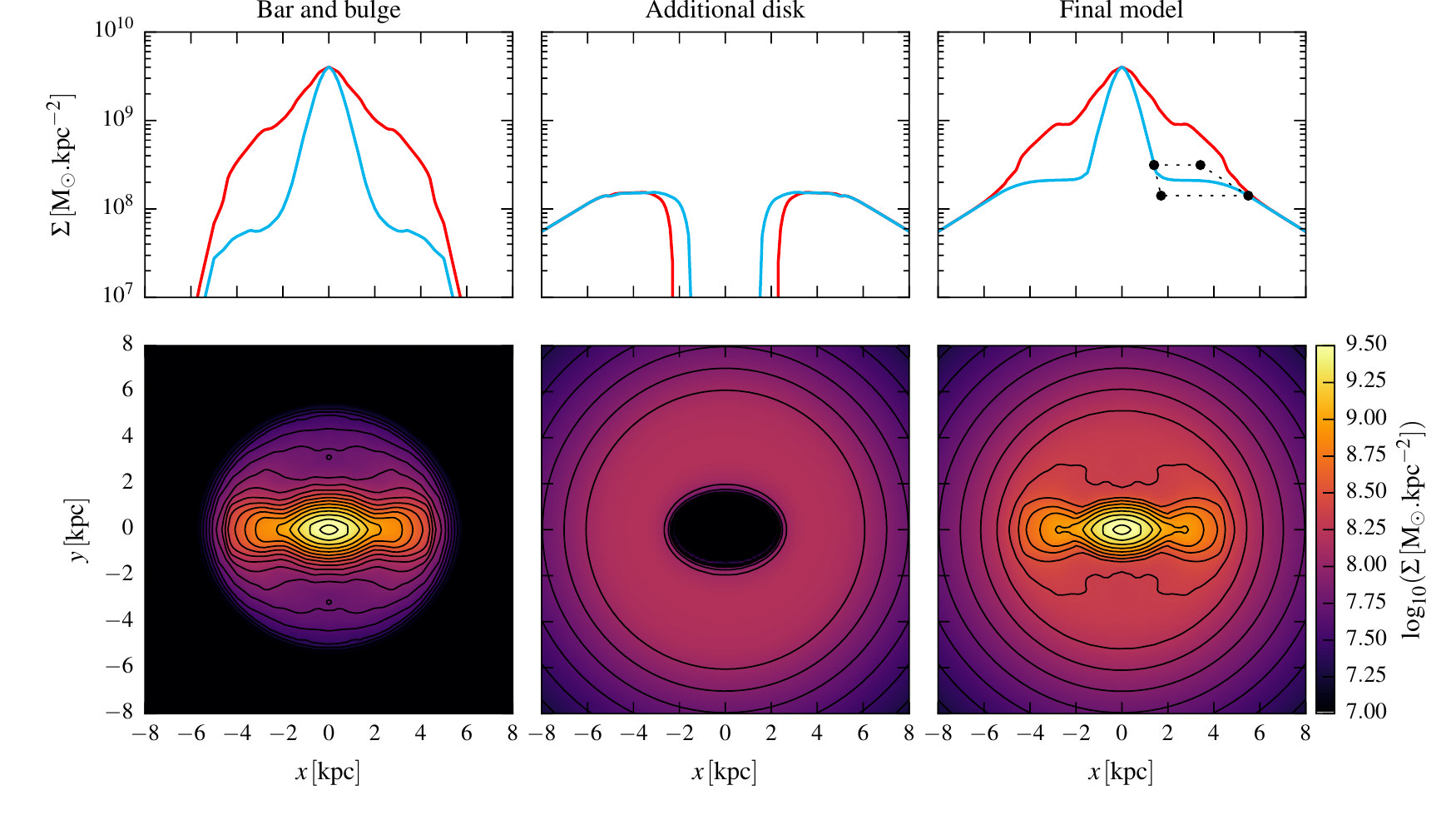}\\
  \caption{Top row: surface density profiles of the Milky Way components along the bar major axis (red) and the bar intermediate axis (blue). Bottom row: face-on surface density of the model components. The dashed region shows the convex hull of the B\'{e}zier curve control points used for the interpolation of the intermediate axis surface density profile between $5.5$ and $1.4\kpc$.}
  \label{fig:staticModel}
\end{figure*}

Our prime interest in modelling the disk is to obtain a reasonable disk contribution to the potential in the bar region and disk foreground contamination when observing the bulge and bar. Outside the bar region ($R\geq5.5\kpc$) we adopt an axisymmetric stellar disk structure with scalelength $h_{R,*}$ and exponential scaleheight $h_{Z,*}$. From papers based on infrared data \citet{BlandHawthorn2016} concluded that $h_{R,*} = 2.6\pm 0.5 \kpc$ with the shorter disk scalelengths in the range $2.1-2.6\kpc$ usually favored by dynamical studies of stellar kinematics or microlensing optical depth towards the bulge \citep{Wegg2016}. We adopt here a fiducial scalelength of $h_{R,*} = 2.4\kpc$ and scaleheight of $h_{Z,*} = 300\pc$ \citep{Juric2008} and test in \autoref{section:stellarAndDarkMatterMass} the effect of a shorter scalelength of $2.15\kpc$ \citep{Bovy2013} and $2.6\kpc$ \citep{Juric2008}.

Following \citet{Bovy2013}, we add the interstellar medium contribution to the potential by modelling it as an additional thin disk with scalelength $h_{R,\rm{ism}} = 2 \times h_{R,*}$ and scaleheight $h_{R,\rm{ism}} = 130\pc$. The disks are normalized to a baryonic local surface density inside $1.1\kpc$ above and below the plane of $\Sigma_{1.1}(R_0) = 51\, \Msun.pc^{-2}$ among which $38\, \Msun.pc^{-2}$ are stars and $13\, \Msun.pc^{-2}$ are interstellar medium \citep{Bovy2013}. 

Inside the bar region, very little is known about the structure of the disk component that surrounds the bar and bulge. We can fortunately constrain this inner disk by combining our knowledge of the bulge and of the disk at the boundary of the bar region (i.e. $5.5\kpc$). In the bulge region, approximated as the interior of an ellipse reaching $2.2\kpc$ along the bar major axis and $1.4\kpc$ along the intermediate axis, the `bulge' model already represents the entire stellar density. Hence the total surface density along the intermediate axis has to smoothly transition between its value at $5.5\kpc$ and the bulge value at $1.4\kpc$. We construct the inner disk in the bar region by first interpolating the total surface density along the intermediate axis between $5.5$ and $1.4\kpc$ using the logarithm of a quadratic B\'{e}zier curve interpolation, whose control points are defined to ensure continuity of the derivative. The disk surface density is then constructed assuming a linear decrease of the ellipticity of the disk isocontours between the bulge and the boundary of the bar region. This procedure is shown in \autoref{fig:staticModel} where the bar and bulge model (left) and its additional disk component (centre) sum up to a total density that smoothly joins the bulge to the disk at the end of the bar region. Although several assumptions enter in joining the bulge, bar and disk as described above, it results in a reasonable global density model for the Milky Way that suits well our purpose of tailoring the initial conditions for the M2M modelling as already stated at the beginning of this section.

\subsection{Dark matter halo}
\label{section:haloModel}

By adding up the bulge, bar and disk as described above, we obtain a 3D density model of the baryonic mass in the Milky Way.
\begin{figure}
  \centering
  \includegraphics{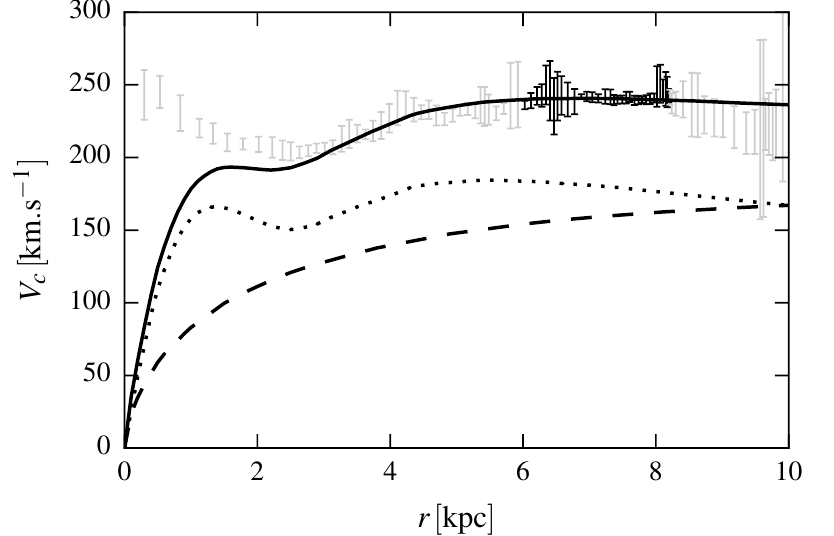}\\
  \caption{Rotation curve of the initial density model of the Milky Way compared to the composite rotation curve of \citet{Sofue2009}. Solid, dotted and dashed lines show respectively the total, baryonic and dark matter rotation curves of our model. Data points used for fitting the dark matter halo as described in the text are shown in black.}
  \label{fig:staticModelRotationCurve}
\end{figure}
\autoref{fig:staticModelRotationCurve} shows the rotation curve $V_c(r)$ of these baryonic mass models together with the composite rotation curve from \citet{Sofue2009} rescaled to a distance to the GC and a local circular velocity of $(R_0, V_0) = (8.2\kpc, 238\kms)$ as described in \autoref{section:InitialConditions}. This baryonic model is insufficient to match the rotation curve of the Milky Way, and we therefore require dark matter within the framework of Newtonian dynamics. Recent studies of dark matter simulations such as \citet{Navarro2010} showed that the innermost regions of dark matter halos were better represented by the Einasto density profile \citep{Einasto1965} than by the NFW profile. Inspired by this, we adopt the three-parameter Einasto density profile given by
\begin{equation}
 \rho_{\rm{DM}}(m) = \rho_0 \, {\rm exp} \left\{-\left(\frac{2}{\alpha} \right) \left[ \left(\frac{m}{m_0}\right)^{\alpha} - 1\right] \right\}
\end{equation}
where $m = \sqrt{x^2 + y^2 + (z/q)^2}$ is the elliptical radius for an assumed vertical flattening of $q=0.8$ \citep{Piffl2014}. We can constrain the halo parameters using the rotation curve but caution has to be taken inside the bar region. The data from \citet{Sofue2009} shown in \autoref{fig:staticModelRotationCurve} are a combination of different data sets and rely mostly on the tangent-point method from terminal velocity measurement of CO and $\rm{H_{I}}$ gas inside the solar circle. Because of the influence of the bar on the gas flows, the tangent point method is likely to be flawed inside the bar region \citep{Englmaier1999, Chemin2015} so we exclude data points inside $6\kpc$ from the GC. We also exclude all data points outside the solar circle, as the rescaling to our assumptions for $(R_0, V_0)$ would require us to take into account the nature of the different data sets entering the work of \citet{Sofue2009}. The rotation curve between $6\kpc$ and $R_0$ provides a good constraint on the average value and slope of the dark matter circular velocity at the solar position, but is not sufficient to constrain the dark matter in the inner region.
The determination of the dark matter contribution in the inner Galaxy requires proper dynamical modelling and is addressed in detail in \autoref{section:patternSpeedAndDarkHalo}. At this stage we assume a dark matter mass inside $2\kpc$ of $0.5\times 10^{10}\Msun$, resulting in the rotation curve shown in \autoref{fig:staticModelRotationCurve}, consistent with our bulge models in \hyperlink{P15}{P15}. In \autoref{section:automaticDMhalo}, we relax the constraint on the dark matter mass inside $2\kpc$ during the modelling process and adapt it directly to match the bulge kinematics. 

A summary of the diverse parameters of our fiducial model for tailoring the initial conditions is given in \autoref{table:ModelParameters}.

\begin{sidewaystable*}
  \caption{Parameters of the Galaxy used to tailor initial conditions.}
  \centering
  \begin{tabular}{lllll}
    \hline
    & Parameter & Fiducial value & Reference & Section\\
    \hline
    Geometry & Distance to GC $R_0$ & $8.2\kpc$ & \citet{BlandHawthorn2016} & \autoref{section:staticMWModels}\\
	     & Bar angle & $28\degree$ & Average between \citet{Wegg2013} and \hyperlink{W15}{W15}  & \autoref{section:staticMWModels}\\
    \hline
    Bulge & 3D density & Fitted M85 & \hyperlink{P15}{P15} & \autoref{section:bulgeAndBarModel} \\
	  & Mass-to-clump ratio $\rm{M/n_{RCG}}$ & 1000 & Direct measurement, Kroupa + Calamida IMF & \autoref{section:empiricalMassToClump}\\
    \hline
    Bar & Thin component& Analytical density  & \hyperlink{W15}{W15} & \autoref{section:bulgeAndBarModel}\\
	& Superthin component& Analytical density  & \hyperlink{W15}{W15} & \autoref{section:bulgeAndBarModel}\\
	& $\rm{M/n_{RCG}}$ of thin bar & Same as bulge & -- & \autoref{section:bulgeAndBarModel}\\
	& $\rm{M/n_{RCG}}$ of superthin bar & 600 & Stellar population models & \autoref{section:bulgeAndBarModel}\\
    \hline
    Stellar disk & Scalelength $h_{R,*}$ & $2.4\kpc$, middle of range $2.15-2.6\kpc$ & \citet{BlandHawthorn2016} & \autoref{section:diskModel}\\
		 & Scaleheight $h_{Z,*}$ & $300\pc$ & \citet{Juric2008} & \autoref{section:diskModel}\\
		 & Local surface density $\Sigma_{1.1, *}(R_0)$ & $38\Msun.\pc^{-2}$& \citet{Bovy2013} &\autoref{section:diskModel}\\
    \hline
    ISM disk & Scalelength $h_{R,\rm{ism}}$ & $2 \times h_{R,*}$ &  \citet{Bovy2013}  & \autoref{section:diskModel}\\
	     & Scaleheight $h_{Z,\rm{ism}}$ & $130\pc$ &  \citet{Bovy2013} & \autoref{section:diskModel}\\
	     & Local surface density $\Sigma_{1.1, \rm{ism}}(R_0)$ & $13\Msun.\pc^{-2}$&  \citet{Bovy2013}  & \autoref{section:diskModel}\\
    \hline 
    Dark matter halo & Profile & Best-fitting Einasto & -- & \autoref{section:haloModel}\\
		     & Flattening & $0.8$ & \citet{Piffl2014} & \autoref{section:haloModel}\\
		     & Outer constraints & $V_c(6\kpc\leq r\leq R_0\kpc)$  & \citet{Sofue2009} & \autoref{section:haloModel}\\
		     & Inner constraint & $M(<2\kpc) = 0.5 \times 10^{10} \, \Msun$ & \hyperlink{P15}{P15} & \autoref{section:haloModel}\\
    \hline
    Dynamical parameters & LSR circular velocity $V_0$ & $238\kms$ & \citet{Schonrich2012}, \citet{Reid2014} & \autoref{section:haloModel}\\
			 & Solar motion in the LSR $(U, V, W)$ & $(11.1, 12.24, 7.25)\kms$ &  \citet{Schonrich2010} & \autoref{section:staticMWModels}\\
			 & Bar pattern speed $\Omega_{b}$ & Systematic search in the range $25-50\kmskpc$ & -- & \autoref{section:staticMWModels}\\
    \hline
  \end{tabular} 
  \label{table:ModelParameters}
\end{sidewaystable*}

\section{Tailoring initial models for modelling the Milky Way}
\label{section:tailoringInitialConditions}
In this section, we use a variant of the M2M method to create a family of Milky Way models with a specified mass distribution and different bar pattern speeds. We adiabatically adapt M85 to the density model of the Galaxy described above and change slowly the bar pattern speed, hence gaining full control on the effective potential of the model.

\subsection{Adiabatic adaptation of the initial conditions}
\label{section:adiabaticEvolution}

We first evaluate the initial stellar and dark matter mass distribution of M85 on respectively a Cartesian grid of $\pm 12\kpc \times \pm 12\kpc \times \pm 2\kpc$ and a radial grid extending to $40\kpc$ with flattening $q=0.8$. We then integrate our fiducial stellar and dark matter density target from \autoref{section:staticMWModels} in the grid cells and define $25$ intermediate targets, log-spaced between the initial model and the target mass distribution. We modify the bar pattern speed by defining $25$ intermediate corotation radii, linearly spaced between the initial model corotation and the target corotation.

The adiabatic adaptation of M85 to the target model then consists of $25$ iterations of the following procedure:
\begin{itemize}
 \item[(i)] Perform an M2M fit of the stellar and dark matter mass distributions to the current target mass distribution while evolving the model in the current rotating potential at constant pattern speed. We run this M2M fit for one time unit, corresponding to the period of one circular orbit at $4\kpc$.
 \item[(ii)] Update the potential to the new particle masses and adapt the pattern speed at which the potential rotates to place the corotation at the next intermediate target value.
 \item[(iii)] Multiply all particle velocities by a factor $\sqrt{\vec{r}\cdot\vec{\nabla}\Phi_{\rm new}} / \sqrt{\vec{r}\cdot\vec{\nabla}\Phi_{\rm old}}$ where $\vec{r}$ is the particle position and $\Phi_{\rm old}$ and $\Phi_{\rm new}$ are the potential respectively before and after the potential update. This step is necessary as the circular velocity in the new potential is different from that of the original model.
\end{itemize}

Each M2M fit to an intermediate target mass distribution is performed using \autoref{equation:FOC1} with the following profit function
\begin{equation}
 F = - \frac{1}{2} \sum_{k,j} (\Delta_j)^2
 \label{equation:profitAdiabatic}
\end{equation}
where $\Delta_j$ is the difference between the model mass and the target mass in cell $j$. Since model observables can be noisy in regions of space where the particle density is low, we use temporal smoothing by replacing the instantaneous model observable $y(t)$ at time $t$ by its temporally smoothed value $\tilde{y}(t)$ defined as
\begin{equation}
\label{equation:temporalSmoothing}
\tilde{y}(t) = \int y(t-\tau)e^{-\alpha \tau} \, d\tau~.
\end{equation}
where $1/\alpha$ is the temporal smoothing time-scale.

We then continue the M2M fit to the final target mass distribution and corotation radius for another $25$ time units. The M2M fit can lead to a broad distribution of weights that has the undesired effect of lowering the model resolution and potentially introducing clumps in the potential due to very massive particles. We added to \nmagic the particle resampling algorithm described in \citet{Dehnen2009}. This algorithm consists of creating a new particle model with equal-weights particles by resampling the original set of particles with a probability proportional to their weights. When multiple selections of a given particle occur, we evolve the parent particle for one orbital time (estimated as $T \sim 2\pi \sqrt{\frac{r}{f_r}}$ where $r$ and $f_r$ are the radius and radial acceleration of the particle) and select multiple particles along the trajectory in phase-space sampled by the parent particle. 

\begin{figure*}
  \centering
  \includegraphics{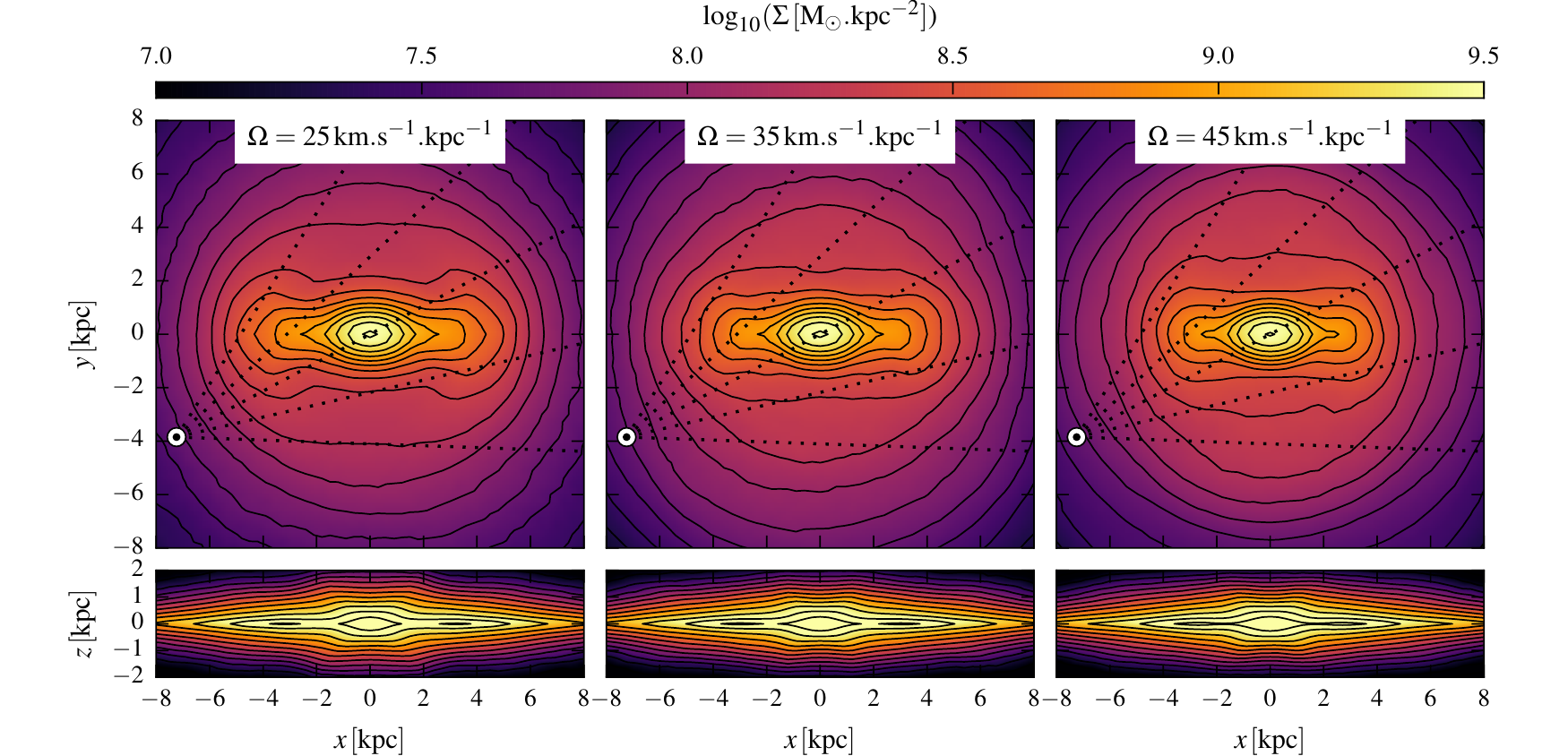}\\
  \caption{Face-on (top row) and side-on (bottom row) surface density of three N-body models adiabatically adapted to the Milky-Way density model of \autoref{section:staticMWModels} for pattern speeds of $25$, $35$ and $45\kms$.}
  \label{fig:adaptedModels}
\end{figure*}

Following this procedure, we create a set of N-body models that broadly matches the static density model of \autoref{section:staticMWModels} with bar pattern speeds in the range $25-50 \kmskpc$. All models have $10^6$ equal weight stellar particles and $10^6$ dark matter particles. \autoref{fig:adaptedModels} shows the surface densities of three N-body models with pattern speeds of $25$, $35$ and $45\kmskpc$. As expected, different pattern speeds lead to slightly different bar shapes but the global mass distribution of the target density is anyway well reproduced in these adiabatically adapted N-body models. 

\subsection{Integration and potential solver}
\label{section:IntegrationAndPotentialSolver}
The particle model is integrated using a drift-kick-drift adaptive leap-frog algorithm in the full gravitational potential rotating at a constant patten speed. The gravitational potential is computed directly from the particle mass distributions using the hybrid grids method described in the appendix of \citet{Sellwood2003}. This hybrid method combines a grid based potential solver on a flat cylindrical grid to evaluate the disk potential with a spherical harmonics potential solver on a spherical grid to evaluate the potential of the dark matter halo. For the cylindrical potential solver, we use the 3D polar grid code from \citet{Sellwood1997}, in a cylindrical grid extending to $12\kpc$ in radius and $\pm 2 \kpc$ in the vertical direction. As a spherical solver, we use the spherical harmonics solver of \citet{DeLorenzi2007} up to order $8$ on a spherical grid extending to $40 \kpc$. 

In addition to the hybrid grid method, we modified the 3D polar grid code in order to allow the resolution of strong vertical gradients. In the original code from \citet{Sellwood1997}, the particle mass distribution is softened using a spherical cubic spline density kernel where the softening scale should not be smaller than the planar scale of the grid cells in order to give accurate results. To keep the number of planar cells under control and still resolve strong vertical gradient we replace the spherical softening by an oblate softening with vertical axial ratios of $0.2$. In the end, we define the grid parameters in order to obtain a planar resolution of $100\pc$ and a vertical resolution of $\sim20\pc$.

\section{\nmagic data constraints for the galactic bulge, bar and disk}
\label{section:nmagicConstraints}
\begin{figure*}
  \centering
  \includegraphics{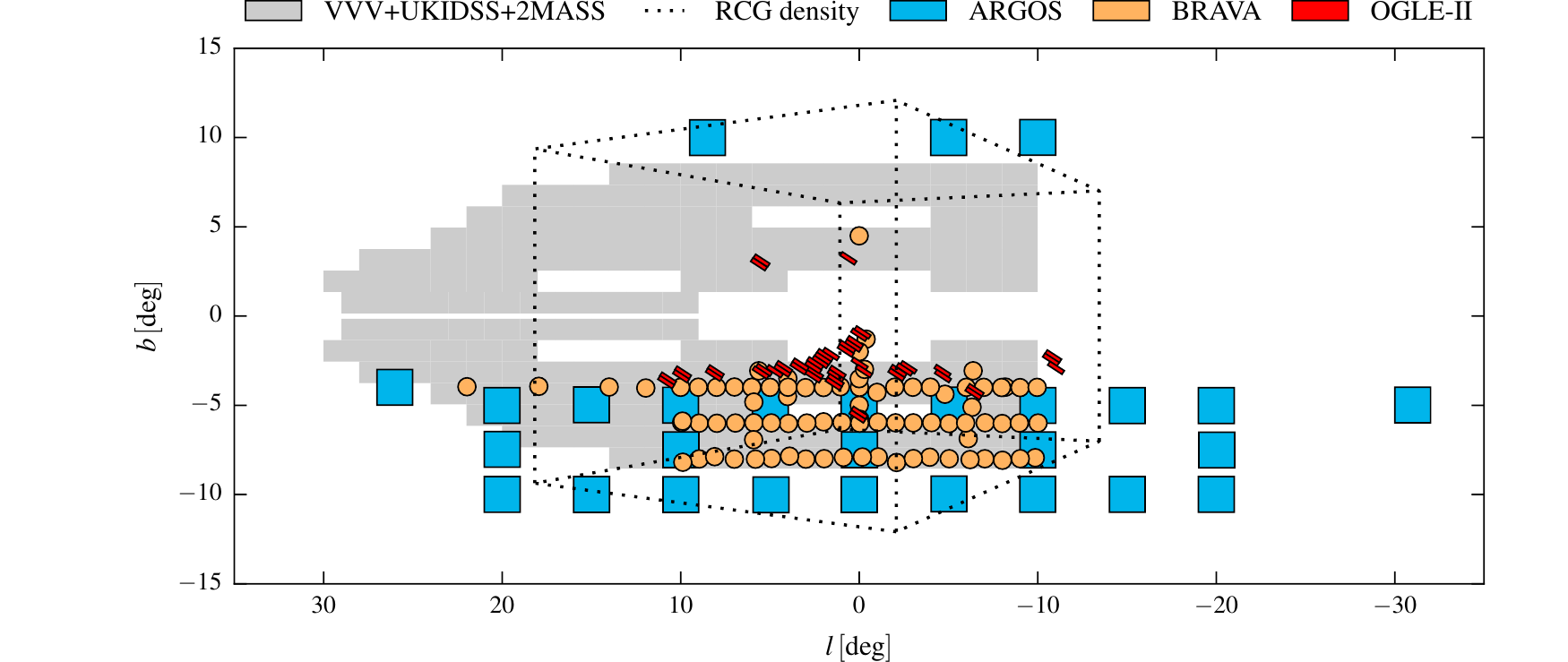}\\
  \caption{Spatial coverage in galactic coordinates of the different data sets used to constrain the models: \argos kinematics (\autoref{section:argos}),\vvv + \ukidss \twomass star counts (\autoref{section:starCounts}), bulge RCG density (\autoref{section:3Ddensity}), \brava kinematics (\autoref{section:bravaKinematics}) and \ogle proper motions (\autoref{section:oglepm}).}
  \label{fig:fieldsPosition}
\end{figure*}

In the previous section, we built a set of initial N-body models with different pattern speeds that already broadly matches the Milky-Way bulge, bar, and disk density. In this section, we describe the different data sets to which we fit our N-body models in \autoref{section:M2Mfitting}. For each data set $k$ and observable $j$, we describe the \nmagic kernels $K_j^k$ to be used in \autoref{equation:observable} for applying observational selection bias to the particles when observing the N-body models. An overview of the spatial coverage of the different datasets described in this section is plotted in \autoref{fig:fieldsPosition}.

\subsection{Density and kinematics of the inner Galaxy from the \argos survey}
\label{section:argos}
The Abundance and Radial velocity Galactic Origin Survey (\argos) is a large spectroscopic survey of about $28000$ stars of the galactic bulge and inner disc. It was designed to sample RCGs all the way from the near disk ($\sim 4.5\kpc$ from the Sun) to the far side of the GC ($\sim 13\kpc$ from the Sun). From the medium-resolution spectra, \citet{Ness2013a} estimated various stellar parameters including radial velocity, stellar temperature and surface gravity. These parameters together with the intrinsically narrow luminosity function of RCGs allow the determination of relatively accurate distances. Hence, the \argos survey provides structural information about the inner disk, bulge and bar together with the radial velocity field in a wide spatial range extending up to $|l| = 20\degree$. In \autoref{section:argosSelection}, we briefly review the selection strategy of the \argos survey and compute its selection function. In \autoref{section:argosKinematics}, we determine distances for all \argos stars and compute the mean velocity and velocity dispersion in several distance modulus bins in each field. Finally in \autoref{section:argosKernel} we describe the \nmagic observable kernels that map the survey selection strategy.

\subsubsection{The \argos selection function}
\label{section:argosSelection}

The \argos stars were selected from the \twomass point-source catalogue \citep{Skrutskie2006}, according to a selection procedure fully described in \citet{Freeman2013}. When computing the \argos selection function, the three following points need to be considered.

\begin{enumerate}
  \item About $1000$ stars are selected for each field. The sampling of a large distance range from $\sim 4.5$ to $\sim 13\kpc$ is achieved by selecting randomly $\sim330$ stars in three magnitude bins defined in the $I$-band ($I \sim 13-14$, $14-15$ and $15-16$).\\
  
  \item The \twomass stars from which the \argos stars are selected in (i) do not correspond to the full point-source \twomass catalogue but only to a high-quality subsample of it. This subsample is defined by a blue color cut $(J-K_s)_0 \geq 0.38$ to remove foreground disk contamination, two magnitude cuts $11.5\leq K_s \leq 14$, and additional criteria to exclude stars with large photometric errors, contamination, blends and low photometric quality. The subsample is biased towards the bright stars for which high-quality imaging is easier to achieve.\\
  
  \item The \twomass catalogue is rapidly incomplete in crowded fields. This affects mostly the three fields at $(l,b) = (0\degree, -5\degree)$ and $(\pm5\degree, -5\degree)$ where \twomass is only $\sim45\%$ complete at $K_s\sim 14$.
\end{enumerate}

We evaluate the incompleteness of the \twomass survey by comparing with the deeper \vvv survey where available \citep{Saito2012}, completeness corrected by \citet{Wegg2013}. In all the \argos fields at the edges of the \vvv coverage, \twomass and \vvv do not deviate significantly from each other over the magnitude range $11.5\leq K_s \leq 14$. Given that the effect of crowding decreases with increasing $|l|$ and $|b|$, we consider that \twomass is complete in all the \argos fields out of the \vvv coverage area.

Extinction is evaluated on a star-by-star basis using the Rayleigh-Jeans Color Excess method (\citealt{Majewski2011}; \hyperlink{W15}{W15}) given by:
\begin{equation}
 \label{equation:extinction}
 A_{K_s} = \frac{A_{K_s}}{E(J-K_s)} [(J-K_s) - (J-K_s)_{\rm RCG})]
\end{equation}
where $(J-K_s)_{\rm RCG}$ is the colors of RCGs and $\frac{A_{K_s}}{E(J-K_s)}$ is a constant that depends on the extinction law. For consistency with \citet{Wegg2013}, we adopt $(J-K_s)_{\rm RCG} = 0.674$ \citep{Gonzalez2011} and $\frac{A_{K_s}}{E(J-K_s)} = 0.528$ \citep{Nishiyama2006}. 

\begin{figure}
  \centering
  \includegraphics{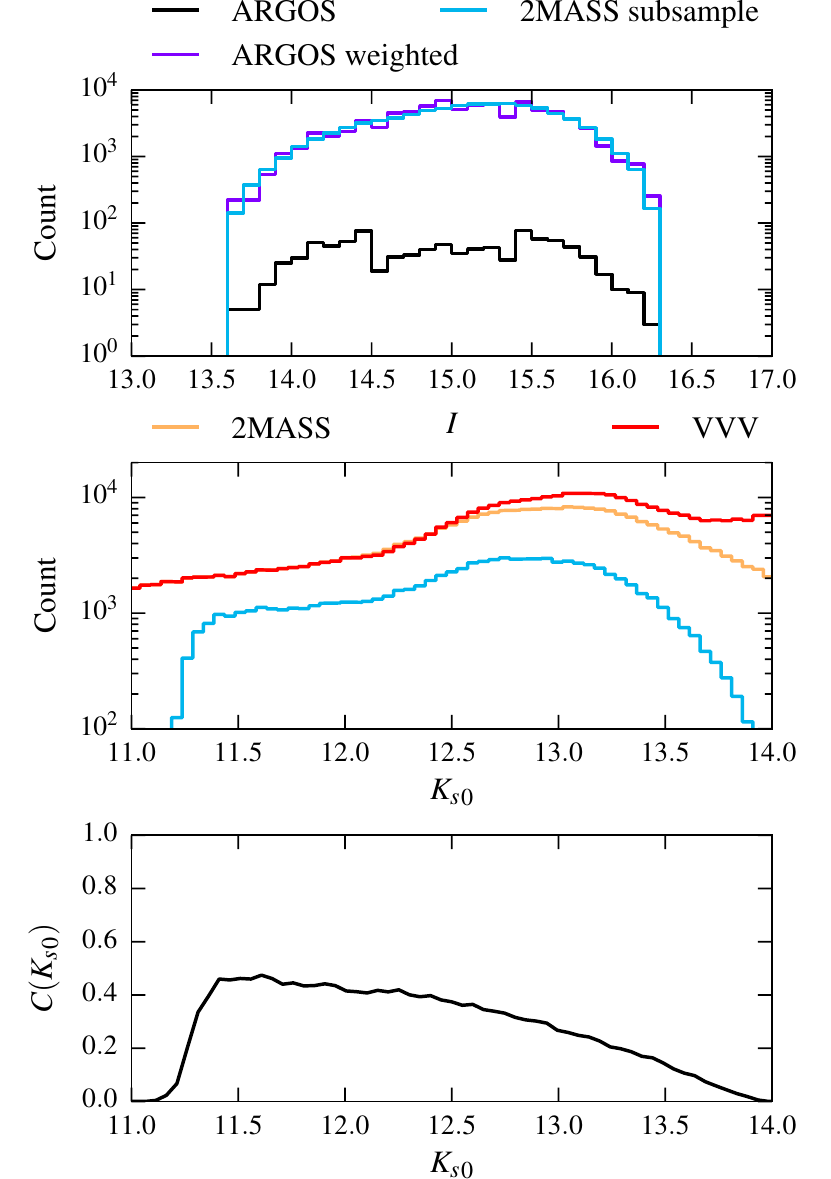}\\
  \caption{Illustration of the \argos selection procedure for the most problematic field at $(l,b) = (0\degree, -5\degree)$. The top panel shows the $I$-band magnitude distribution for the \argos stars (black) and the \twomass data subsample of good photometric quality (blue), in good agreement with the weighted \argos stars (purple). The middle panel shows the extinction-corrected magnitude distributions of the \twomass catalogue (yellow), the completeness-corrected \vvv data (red) and the \twomass subsample of good photometric quality considered by the \argos team (blue). The bottom panel shows the final selection function $C({K_s}_0)$ for this field.}
  \label{fig:argosSelectionFunction}
\end{figure}

To take points (ii) and (iii) into account, we construct the selection function $C({K_s}_0)$ that gives the probability for a star of extinction-corrected magnitude ${K_s}_0$ to belong to the high-quality photometric \twomass subsample from which the \argos stars were selected. This is evaluated empirically as shown in \autoref{fig:argosSelectionFunction}. We finally define $C({K_s}_0)$ as the fraction of stars with magnitude ${K_s}_0$ from the completeness-corrected \twomass catalogue that are also in the high-quality \twomass subsample.

To correct for the selection bias (i) we assign a weight $w_k$ to each of the \argos stars, corresponding to the fraction of selected stars out of the number of stars from the \twomass subsample of good photometric quality present in the considered $I$-band bin.

\autoref{fig:argosSelectionFunction} illustrates this selection procedure for the field at $(l,b) = (0\degree, -5\degree)$, the field most affected by selection effects.

\subsubsection{Density and kinematics as a function of distance}
\label{section:argosKinematics}

Assuming that all \argos stars are RCGs and using the star-by-star extinction of \autoref{equation:extinction}, we estimate the distance moduli $\mu_{K_s}$ of all \argos stars as 
\begin{equation}
 \label{equation:distanceModulus}
 \mu_{K_s} = K_s - A_{K_s} - M_{K_s, \rm{RCG}}
\end{equation}
where we adopt an absolute magnitude of RCGs of $M_{K_s, \rm{RCG}}=-1.72$ for consistency with \citet{Wegg2013}.

Stars in the \argos sample are either real RCGs, for which distances are accurately determined by \autoref{equation:distanceModulus}, or red giants that happen to be at the right distance to appear with similar color and magnitude as the real RCGs. For these giants, the distance estimate given by \autoref{equation:distanceModulus} can be wrong by several magnitudes. To minimize their effect, we follow \citet{Ness2013a} and evaluate the absolute magnitude $M_{K_s}$ of every star using the surface gravity $\log g$ measurement obtained from fitting the spectra together with the PARSEC isochrones \citep{Bressan2012, Chen2014, Tang2014} and assuming a $10\Gyr$ old population, a Kroupa IMF and the overall metallicity distribution of all the \argos stars. We then statistically remove non-RCG stars by replacing the weights $w_k$ of all \argos stars with $w_k \times \omega(M_{K_s})$, where $\omega$ is a weighting function depending on the inferred absolute magnitude. The uncertainty in $\log g$ is $\sim 0.3\, \rm{mag}$, which is equivalent to an uncertainty in $M_{K_s}$ of $0.7 \, \rm{mag}$. Original work from the \argos team chose for $\omega$ a top-hat function around $M_{K_s, \rm{RCG}}$, identifying in this way what they call the `probable RCGs'. In order to take advantage of the full sample of stars, we adopt instead the weighting $\omega(M_{K_s}) = G_{[M_{K_s, \rm{RCG}}, 0.7]}(M_{K_s})$  where  $G_{[\mu,\sigma]}$ is the Gaussian function of mean $\mu$ and standard deviation $\sigma$.

Finally, we bin the distance modulus space in bins of $0.25 \, \rm{mag}$ and compute for each field the number of stars $n_{j,m}$, the mean radial velocity $\vel_{j,m}$ and radial velocity dispersion $\sigma_{j,m}$ of the stars in field $j$ and distance modulus bin $m$, taking into account each star weight $w_k$. Errors in those quantities are computed by $1000$ bootstrap resamplings.

\subsubsection{\nmagic observable for the \argos data}
\label{section:argosKernel}
In order to map the observational criteria to observing an N-body model, we first need to study the stellar population that falls into the \argos sample in more detail. Using the PARSEC isochrones for a $10\Gyr$ old population with a Kroupa IMF and the overall metallicity distribution of all the \argos stars, we predict the luminosity function $\Phi$ of the stars that matches the \argos color and magnitude cuts. The RCGs then appear as a sharp peak over a large background of red giants, located at $M_{K_s, \rm{RCG}} = -1.47$, slightly fainter than the assumed value of $M_{K_s, \rm{RCG}} = -1.72$ used by \citet{Wegg2013} for measuring the 3D density of the galactic bulge. For internal consistency between our data sets we shift the isochrone luminosity function to agree with the maximum of the RCG peak at $M_{K_s, \rm{RCG}} = -1.72$.

Let us now observe an N-body model by effectively turning the particles into stars and applying the selection procedure of the \argos survey. We consider a particle at distance modulus $\mu_i$ and note $f_i(\mu_{K_s})$ the distribution of observed distances of mock stars drawn from $\Phi(K_s)$ at the position of the considered particle when assuming that they are all RCGs. A mock star with an absolute magnitude $M_{K_s}$ will be inferred to lie at a distance $\mu_{K_s} = M_{K_s} + \mu_i - M_{K_s, \rm{RCG}}$. $f_i(\mu_{K_s})$ can be written as

\begin{equation}
  \label{equation:distanceDistributionOfAParticle}
  \begin{split}
 f_i(\mu_{K_s}) = &\Phi(\mu_{K_s} - \mu_i + M_{K_s, \rm{RCG}}) \times \overline{\omega}(\mu_{K_s} - \mu_i + M_{K_s, \rm{RCG}})\\
  & \times C(\mu_{K_s}+M_{K_s, \rm{RCG}})
  \end{split}
\end{equation}
where $C(K_{s0})$ is the selection function computed in \autoref{section:argosSelection}. The mean weight $\overline{\omega}(M_{K_s})$, average of the weights obtained after statistical removal of non-RCGs from a mock measurement of $M_{K_s}$, with accuracy $0.7 \, \rm{mag}$ is given by
\begin{equation}
 \overline{\omega}(M_{K_s})  = \int \omega(M)\times G_{[M_{K_s},0.7]}(M) \, dM.
\end{equation}

\begin{figure}
  \centering
  \includegraphics{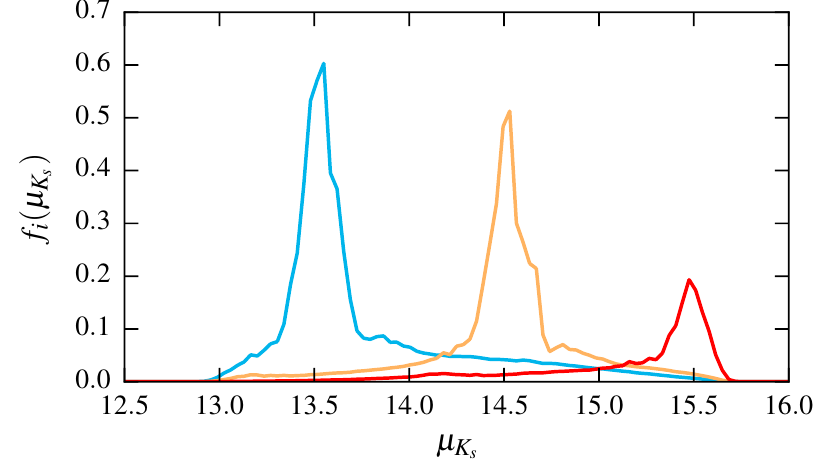}\\
  \caption{Distribution of observed distances $f_i(\mu_{K_s})$ of mock stars drawn from a particle at distance modulus $\mu_i = 13.5$ (blue), $\mu_i = 14.5$ (yellow) and $\mu_i = 15.5$ (red), in the field at $l,b = 10\degree, -5\degree$. The  background of giants present in the luminosity function is attenuated here by the weighting $\overline{\omega}(\mu_{K_s} - \mu_i + M_{K_s, \rm{RCG}})$ together with the incompleteness of the \twomass subsample for faint magnitudes.}
  \label{fig:distanceDistributionOfAParticle}
\end{figure}

The distribution of distances $f_i(\mu_{K_s})$ resulting from \autoref{equation:distanceDistributionOfAParticle} is shown in \autoref{fig:distanceDistributionOfAParticle} for three particles at at $\mu_i = 13.5$, $14.5$  and $15.5$. The distribution is narrow in all cases, showing that the \argos survey provides accurate distances. The spreading in distances due to the background of giants is minimized thanks to the selection bias towards nearby stars and also the extra information provided by the measurement of $\log g $.

As \nmagic observables we adopt the number count and first and second mass-weighted velocity moments. The corresponding kernels for a field $j$ and distance modulus $m$ to be used in \autoref{equation:observable} are given by
\begin{equation}
 K_{j, m}^{\rm{\argos}, 0} = \delta_{j,m}^{\rm{\argos}} (\vec{z}_i)
\end{equation}
\begin{equation}
 K_{j, m}^{\rm{\argos}, 1} = \delta_{j,m}^{\rm{\argos}} (\vec{z}_i) \times \frac{\vel_i}{W_{j,m}^{\rm{\argos}}}
\end{equation}
and 
\begin{equation}
 K_{j, m}^{\rm{\argos}, 2} = \delta_{j,m}^{\rm{\argos}} (\vec{z}_i) \times \frac{\vel_i^2}{W_{j,m}^{\rm{\argos}}}
\end{equation}
where $\vel_i$ is the radial velocity of particle $i$, $W_{j,m}^{\rm{\argos}}$ is given by
\begin{equation}
 W_{j,m}^{\rm{\argos}} = \sum_i w_i \delta_{j,m}^{\rm{\argos}}(\vec{z}_i)
\end{equation}
and $\delta_{j,m}^{\rm{\argos}}$ by
\begin{equation}
\label{equation:deltaArgos}
 \delta_{j,m}^{\rm{\argos}} (\vec{z}_i) = 
 \begin{cases}
   \int_{\mu(m)}^{\mu(m+1)} f_i(\mu)\, d\mu  & \text{if $i \in $ field $j$, }\\
   0 & \text{otherwise}
  \end{cases}
\end{equation}
with $\mu(m)$ and $\mu(m+1)$ the boundaries of the distance modulus bin $m$.

\subsection{Magnitude distribution of the bulge and bar}
\label{section:starCounts}
In the region delimited by $|l|\leq40\degree$ and $|b|\leq9\degree$, we use the combined catalogue of the \vvv, \ukidss and \twomass surveys from \hyperlink{W15}{W15}. For each line of sight, this catalogue consists of histograms of distance moduli of stars $\mu_{K_s}$ defined as in \autoref{equation:distanceModulus}. Each of these histograms shows an exponential background distribution of stars plus an overdensity of stars due to the RCGs located in the bulge or bar, as shown in \autoref{fig:MasstoClump} for one bulge field. Information on the density is very hard to extract from the exponential background distribution of stars as it arises from the convolution of the line of sight density with the luminosity function of giant and dwarf stars that is very broad, poorly known and likely to vary from field to field and along the line of sight. Hence, we restrict our use of the histograms to the fields that show a significant excess of stars above the exponential background. To do so, we follow \hyperlink{W15}{W15} and first fit a Gaussian plus an exponential to the distribution of distance moduli, separately in each field. We remove from further considerations all lines of sight where either the RCG bump is not detected to at least three sigma or the exponential background slope is too small, indicating incompleteness (see \hyperlink{W15}{W15}). The spatial coverage of the remaining fields is plotted in \autoref{fig:fieldsPosition}.

The kernels of our model observables in a field $j$ and distance modulus bin $m$ bounded by $\mu(m)$ and $\mu(m+1)$ are given by
\begin{equation}
 K_{j, m}^{\rm{hist}}(\vec{z}_i) = 
\begin{cases}
\frac{1}{\rm{M/n_{RCG}}} \times \int_{\mu(m)}^{\mu(m+1)} \Phi(\mu - \mu_i) \, d\mu & \text{if $i \in$ field $j$,}\\
0 & \text{otherwise}
\end{cases}
\end{equation}
where $\rm{M/n_{RCG}}$ is the mass-to-clump ratio described in \autoref{section:empiricalMassToClump} and $\Phi$ the luminosity function for RCGs only, expressed in distance modulus (see equation 17 in \hyperlink{W15}{W15}). The exponential background of stars, absent from the model observables, needs to be introduced before comparing model to data. To do so, we follow \hyperlink{W15}{W15} and first fit an exponential distribution $y_j^e$ to the difference between the model observable $y_j^{\rm{hist}}$ and full data histogram $Y_j^{\rm{hist}}$. This exponential background of stars is then included in the data-model comparison by replacing  $y_j^{\rm{hist}}$ by $y_j^{\rm{hist}} + y_j^e$ in \autoref{equation:delta}.

For all fields with $|b|\geq1.35\degree$, we assume a $10\Gyr$ old stellar population with a Kroupa IMF and the metallicity distribution of \citet{Zoccali2000} in Baade's window and compute $\Phi$ from the PARSEC isochrones. After removing the exponential background we scale $\Phi$ to our fiducial mass-to-clump ratio of $1000$ (see \autoref{section:empiricalMassToClump}). 

For $|b|<1.35\degree$, as already discussed in \autoref{section:bulgeAndBarModel}, \hyperlink{W15}{W15} found evidence for a superthin bar component, present mostly near the bar end. This superthin bar is likely to be formed by younger stars as indicated by its extremely small scaleheight of only $45\pc$. Detailed modelling of the superthin component is beyond the scope of this paper, but as our goal is to constrain the gravitational potential, its mass has to be included in the modelling. We follow \hyperlink{W15}{W15} and assume a constant star formation rate for the superthin component to compute its luminosity function. The superthin bar population has a mass-to-clump ratio of $600$, lower than the old bar population. For fields with $|b|\leq 1.35\degree$ we use a superposition of the thin bar and superthin bar populations, using the ratio of the densities given by the parametric models of \hyperlink{W15}{W15}. 

For efficiency we combine the original data of \hyperlink{W15}{W15} in cells of $2\degree \times 1.2\degree$ in galactic coordinates for $|b|\geq 1.35\degree$ and cells of $2\degree \times 0.6\degree$ for $|b|<1.35\degree$ and symmetrize with respect to the $b=0\degree$.

\subsection{3D density of the bulge and outer disk}
\label{section:3Ddensity}

In the bulge we constrain the stellar density using the 3D density of RCGs measured by \citet{Wegg2013}, scaled to stellar mass density using our fiducial mass-to-clump ratio. The map covers a box of $(\pm 2.2 \times \pm 1.4 \times \pm 1.2) \kpc$ along the bulge principal axes but is incomplete within $\pm 150\pc$ above and below the galactic plane because of large extinction and crowding. We found in \hyperlink{P15}{P15} that the vertical density profile of our N-body models of B/P bulges was well represented by a vertical $\rm{sech}^2$ profile. Thus, to fill the mid-plane gap in the RCG density map we use the fiducial extrapolation of \hyperlink{P15}{P15}, obtained by fitting a $\rm{sech}^2$ profile to each vertical slice through the bulge. This extrapolation provides us with the full 3D density of what we call the \emph{smooth bulge}.  In \autoref{section:missingCentralMass}, we consider an extra in-plane disk component and show that indeed an in-plane over density is required to match our bulge kinematic data. We finally integrate the RCGs map on a grid of $(30\times28\times32)$ cells, and the corresponding observable kernels $K_j^{\rm{RCG}}$ are given by
\begin{equation}
 K_j^{\rm{RCG}} (\vec{z}_i) = 
 \begin{cases}
   \rm{M/n_{RCG}}^{-1} & \text{if $i \in $ cell $j$,}   \\
   0 & \text{otherwise.}
  \end{cases}
  \label{equation:RCGKernel}
\end{equation}
 
Outside the bar region, for cylindrical radius larger than $5\kpc$, we use the 3D density of the disk of \autoref{section:diskModel} evaluated on a large 3D density grid using a mass-in-cell kernel similarly to \autoref{equation:RCGKernel}.  

\subsection{Bulge kinematics from the \brava survey}
\label{section:bravaKinematics}
The \brava survey is a large spectroscopic survey of about $10000$ M giant stars, mostly towards the bulge \citep{Rich2007, Howard2008, Kunder2012}. We use only the fields with $|l|\leq 10\degree$ as we found in \hyperlink{P15}{P15} that the disk contamination could become significant outside the bulge. The selected \brava fields provide $82$ measurements of the mean radial velocity and velocity dispersion through the bulge. 

As \nmagic observables we use here the first and second weighted velocity moments whose kernels are given by:
\begin{equation}
\label{equation:bravaKernel1}
 K_{j}^{\rm{\brava}, 1}(\vec{z}_i) = \delta_j^{\rm{\brava}}(\vec{z}_i)  \times \frac{\vel_i}{W_j^{\rm{\brava}}}
\end{equation}
and 
\begin{equation}
\label{equation:bravaKernel2}
 K_{j}^{\rm{\brava}, 2}(\vec{z}_i) = \delta_j^{\rm{\brava}}(\vec{z}_i) \times \frac{\vel_i^2}{W_j^{\rm{\brava}}}
\end{equation}
where $\vel_i$ is the radial velocity of particle $i$ and $\delta_j^{\rm{\brava}}$ is the selection function of the \brava survey and $W_j^{\rm{\brava}}$ is given by
\begin{equation}
 W_j^{\rm{\brava}}  = \sum_i w_i \delta_j^{\rm{\brava}}(\vec{z}_i).
\end{equation}
As shown in \hyperlink{P15}{P15} the \brava survey is biased towards nearby stars and the selection function is given by:

\begin{equation}
 \delta_j^{\rm{\brava}} (\vec{z}_i) = 
 \begin{cases}
   r_i^{-1.4} & \text{if $i \in $ field $j$ with $|y_i|<3.5\kpc$,}\\
   0 & \text{otherwise.}
  \end{cases}
\end{equation}

\subsection{Bulge proper motions from the \ogle-II survey}
\label{section:oglepm}

\begin{figure}
  \centering
  \includegraphics{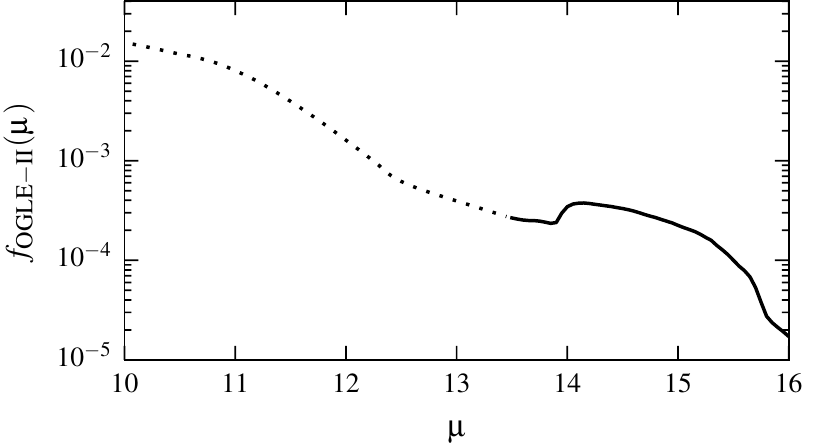}\\
  \caption{Selection function of the \ogle survey as a function of distance modulus for the field at $(l,b) = (1.0\degree, -3.7\degree)$. The solid line indicates the selection function in the range of distances considered in this work. The large contribution of nearby stars indicated by the dashed line is ignored.}
  \label{fig:ogleSelection}
\end{figure}

Proper motion data in the bulge for more than half a million stars have been measured by the \ogle survey. In this paper, we chose to use the \ogle proper motions to compare with model predictions rather than using them in the model fitting. We use the proper motion dispersions of RCG in the bulge from the \ogle-II survey as computed by \citet{Rattenbury2007} from the proper motion catalogue of \citet{Sumi2004}. RCGs are selected in an ellipse of the dereddened $I_0$ versus $(V-I)_0$ CMD centred on the expected locus of the red clump at $I_0 = 14.6$ and $(V-I)_0 = 1.0$. Stars with proper motion error larger than $1\masyr$ in either the $l$ or the $b$ direction were excluded from the sample, as well as stars with total proper motion larger than $10\masyr$ which are likely to belong to the foreground disk. We compute the selection function by using again the PARSEC isochrones for a $10\Gyr$ population, a Kroupa IMF and the metallicity distribution of \argos stars in the bulge. The cut in proper motion error introduces a bias towards nearby stars as the error is more likely to be large for faint stars. To model this effect we first compute in each field $j$ the fraction of stars that pass the error cut threshold as a function of their extinction corrected $I_0$ magnitude, denoted by $C_{\sigma_{\mu}, j}(I_0)$. Then from the isochrones we compute for each distance $\mu$ the distribution of $I_0$ magnitudes $C_j(\mu, I_0)$ of stars that end up inside the ellipse selection region of the $I_0$ versus $(V-I)_0$ diagram. The final selection function $f^{\rm \ogle}(\mu)$ is then given by
\begin{equation}
 f^{\rm{\ogle}}(\mu) = \int C_j(\mu, I_0)\times C_{\sigma_{\mu}, j}(I_0)\, dI_0
\end{equation} 
and is shown in \autoref{fig:ogleSelection} for the field at $(l,b) = (1.0\degree, -3.7\degree)$. The large theoretical contribution of low distance modulus is due to faint main-sequence stars in the disk that are close enough to fall into the ellipse selection of the CMD. As the stellar population of the nearby disk is very different from the old bulge population, the selection function should not be trusted at small distances. We adopt a simple distance cut and discard contribution from any particle at distances less than $5\kpc$. Disk contamination is also likely to be removed in the data thanks to the cut in total proper motion.

Observable kernels are defined similarly to \hyperref[equation:bravaKernel1]{Equations ~\ref{equation:bravaKernel1}} and \hyperref[equation:bravaKernel2]{Equations ~\ref{equation:bravaKernel2}} by replacing the radial velocity by the proper motion in the heliocentric frame, with the selection function $\delta_j^{\rm{\ogle}}$ given by:
\begin{equation}
 \delta_j^{\rm{\ogle}} (\vec{z}_i) = 
 \begin{cases}
   f^{\rm \ogle}(\mu_i) & \text{if $i \in $ field $j$ with $r_i>5\kpc$}\\
    & \text{ and $\sqrt{{\mu^*_l}_i^2 + {\mu^*_b}_i^2} < 10\masyr$,}   \\
   0 & \text{otherwise.}
  \end{cases}
\end{equation}

The errors in proper motion dispersion quoted by \citet{Rattenbury2007} are statistical errors, very small due to the large number of observed stars. However, \citet{Rattenbury2007} noted that adjacent fields could show variations in proper motion dispersion of up to $0.2\masyr$. To take those systematics into account, we replace the error bars by adding in quadrature to statistical error a systematic error of $0.1\masyr$.

\section{Dynamical modelling of the bar region}
\label{section:M2Mfitting}
In this section, we use the M2M method to fit the N-body models constructed in \autoref{section:tailoringInitialConditions} to the data described in \autoref{section:nmagicConstraints}.

\subsection{M2M formalism}
\label{section:M2Mformalism}

Indexing again an observable by $j$ and a data set by $k$, the difference between the model observable $y_j^k(t)$ and the real data $Y_j^k$ is evaluated through the residual $\Delta_j^k(t)$ defined as
\begin{equation}
\label{equation:delta}
\Delta_j^k(t) = \frac{y_j^k(t) - Y_j^k}{\sigma(Y_j^k)}
\end{equation}
where $\sigma(Y_j^k)$ is the error on $Y_j^k(t)$.

Following \citet{DeLorenzi2007}, we match our data adopting the profit function
\begin{equation}
 F = - \frac{1}{2} \sum_{k,j} \lambda_k(\Delta_j^k)^2 + \mu S.
 \label{equation:profit}
\end{equation}
The first term in \autoref{equation:profit} is a weighted chi-square term where the $\lambda_k$ are numerical weights for the different data sets (see \citealt{Long2010}). The second term is an entropy term, forcing the particle weight distribution to remain narrow around some pre-determined prior values, improving hence the convergence of the individual particle weights. We use the pseudo-entropy $S$ of \citet{Morganti2012} given by
\begin{equation}
 S = -\sum_{i=1}^N w_i \, \left [\log \left(\frac{w_i}{\hat{w_i}}\right) - 1\right ]
 \label{equation:entropy}
\end{equation}
where $\hat{w_i}$ are prior weights, chosen to be the mean stellar weight and mean dark matter weight for respectively the stellar and dark matter particles.

Using the observable kernels $K_j^k$ described in \autoref{section:nmagicConstraints}, \autoref{equation:FOC1} becomes
\begin{equation}
  \label{equation:FOC2}
  \begin{split}
  \frac{dw_i}{dt} = - \varepsilon w_i \bigg[ &\mu \log\left(\frac{w_i}{\hat{w_i}}\right) \\
  & + \sum_k \lambda_k \sum_{j} \left ( K_j(\vec{z}_i) + w_i\frac{\partial K_j(\vec{z}_i)}{\partial w_i} \right ) \frac{\Delta_j^k(t)}{\sigma(Y_j^k)} \bigg ]. 
  \end{split}
\end{equation}
When the observable kernels do not depend on the weights, \citet{Syer1996} and \citet{DeLorenzi2007} showed that the observables converge exponentially on a time-scale $O(\varepsilon^{-1})$ provided the initial model is sufficiently close to the solution. We find in practice that weight-dependent kernels, as chosen for the \argos and \brava surveys, can be introduced without affecting the convergence provided (i) the derivative of the kernel with respect to the weight is taken into account in \autoref{equation:FOC2}, (ii) some other observables constrain the absolute value of the weights on the spatial domain where the weight-dependent kernel is non-null. We numerically evaluate the convergence of the weights by following the method of \citet{Long2010} considering that a given particle weight has converged if its maximum relative deviation from its mean value on the previous orbit is smaller than $10\%$.

\subsection{Fitting procedure and parametrization}
\label{section:fittingParameters}

After the adiabatic evolution of the initial model, a typical \nmagic fit consists of three phases:
\begin{enumerate}
 \item Temporal smoothing: evolve and observe the model for a time $\rm{T_{smooth}}$ to initialize the observables.
 \item M2M fit: evolve the model and modify the particle weights according to \autoref{equation:FOC2} for a time $\rm{T_{M2M}}$.
 \item Relaxation: evolve and observe the model for a time $\rm{T_{relax}}$ to test the stability of the fit.
\end{enumerate}

We use $4$, $40$ and $16$ internal time units for respectively $\rm{T_{smooth}}$, $\rm{T_{M2M}}$ and $\rm{T_{relax}}$, with one internal unit of time corresponding to the period of one circular orbit at $4\kpc$ (i.e. about $125\,\rm{Myr}$). We adopt a smoothing time-scale of $1/\alpha = 1$ model time units. The force of change parameter $\varepsilon$ is fixed to $\varepsilon = 10^{-1} \times w_0$ where $w_0$ is the mean particle mass. We determined the $\lambda_k$ of \autoref{equation:profit} by analyzing the contribution to the bracket term of \autoref{equation:FOC2} of each set of observables separately. We found that the values of $\lambda_{\rm density} = 1$ for all density observables and $\lambda_{\rm kinematics} = 10$ for all kinematic observables give approximately the same median strength of the force of change to each set of observables. We thus adopted these values for the $\lambda_k$ and checked that only minimal differences are obtained for values of $\lambda_{\rm kinematics} = 5$ or $20$.
The priors of the entropy term of \autoref{equation:entropy} are fixed to the mean particle weight and the magnitude of the entropy is fixed to $\mu=10^4$. This value allows more than 97\% of the particle to converge while resulting in a weight distribution that extends only to $\pm1 \rm{dex}$ from the priors. With these settings and the observables described above, each \nmagic run typically requires $\sim 190$ CPU core-hour to be completed.

\section{Understanding the bulge dynamics}
\label{section:bulgeDynamics}

In this section, we show how the bar pattern speed, dark matter density and in-plane stellar bulge density influence the bulge kinematics and how we can recover and constrain the effective potential by fitting our models to the data described in \autoref{section:nmagicConstraints}.

\subsection{Signature of the pattern speed and inner dark matter halo in the bulge kinematics}
\label{section:patternSpeedAndDarkHalo}
\begin{figure*}
  \centering
  \includegraphics{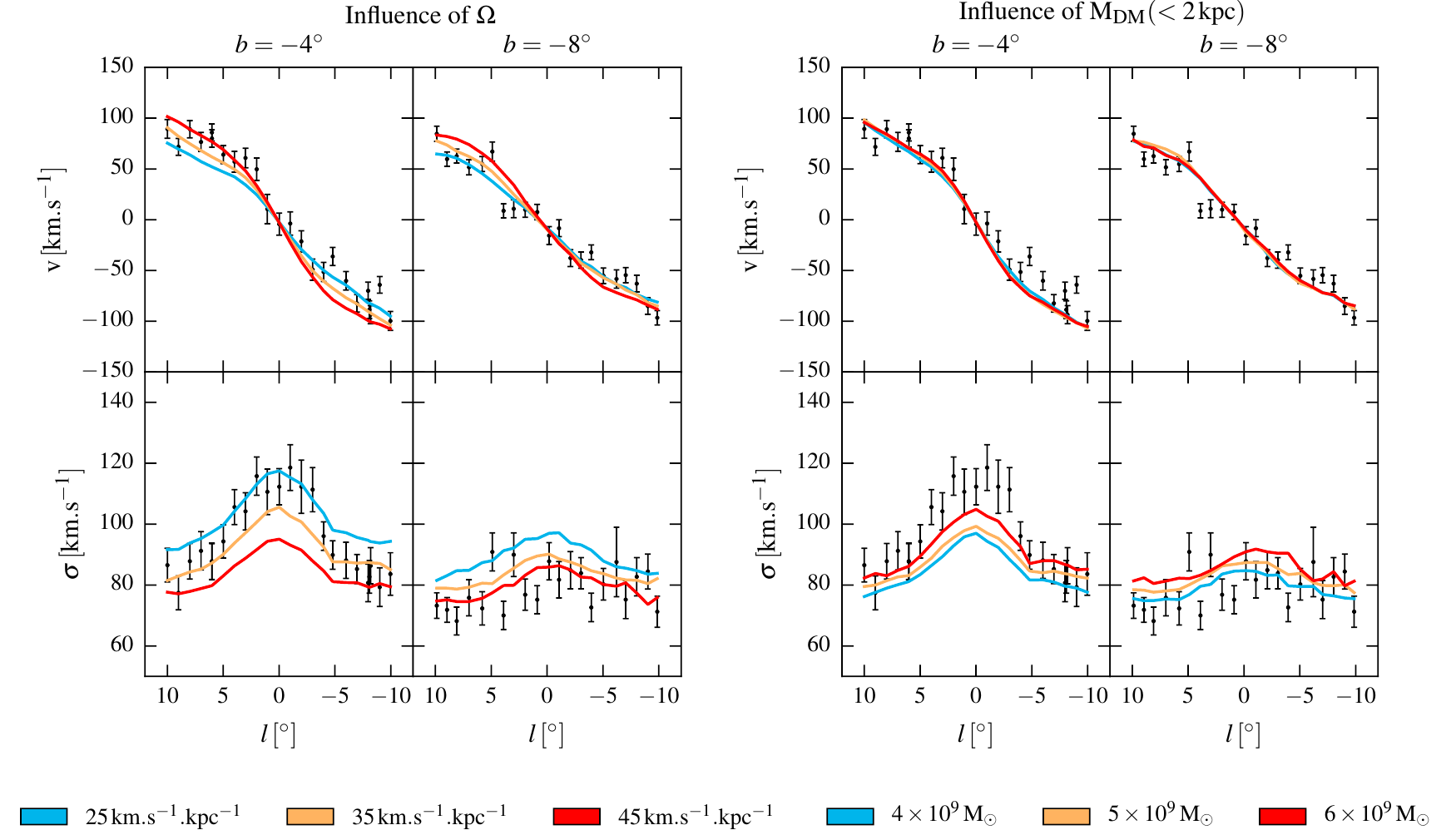}\\
  \caption{Kinematic signature of the pattern speed and dark matter mass in the bulge. The first column shows the influence of the pattern speed on the \brava kinematics for a constant stellar and dark matter mass distribution. The three models plotted here are for pattern speeds of $25\kmskpc$ (blue), $35\kmskpc$ (yellow) and $45\kmskpc$ (red) for a constant bulge dark matter mass of $\rm{M_{DM}(<2\kpc) = 0.5 \times 10^{10} \,  \Msun}$. The pattern speed has a strong influence on both the mean velocity and the velocity dispersion in the bulge. The second column shows the influence of the dark matter mass on the bulge for $0.4 \times 10^{10} \, \Msun$ (blue), $0.5 \times 10^{10} \, \Msun$ (yellow) and $0.6 \times 10^{10} \, \Msun$ (red) and a constant pattern speed of $40\kmskpc$. The dark matter in the bulge has a negligible influence of the mean \brava velocity and can thus be recovered from the velocity dispersion once the pattern speed is fixed.}
  \label{fig:bulgeKinematics}
\end{figure*}

The first column of \autoref{fig:bulgeKinematics} shows the \brava kinematics for three models with different pattern speeds between $25$ and $45\kmskpc$, fitted to our bulge and long bar data in the same dark matter halo. The effect of the bar pattern speed is clearly visible both in the mean velocity and velocity dispersion. Increasing pattern speed leads to an increase in mean velocity but a decrease in the velocity dispersion. The virial theorem provides intuitive explanation of this: for a given mass distribution and hence a given potential energy, a larger pattern speed places more kinetic energy in pattern rotation, leaving less energy available to build up random motions. The second panel of \autoref{fig:bulgeKinematics} shows the \brava kinematics for three models with the same pattern speed and stellar mass density but different dark matter masses in the bulge. As already found in \hyperlink{P15}{P15}, large masses increase the dispersion but leave the mean velocity essentially unchanged. Hence, for a given pattern speed and stellar density model, we can recover the dark matter mass in the bulge from the \brava velocity dispersions. 

\subsection{Recovering the best dark matter halo}
\label{section:automaticDMhalo}

In \autoref{section:haloModel}, we constructed a first-guess dark matter density in the Galaxy by fitting an Einasto profile to the rotation curve between $6\kpc$ and $R_0$ under the constraint of an assumed dark matter mass inside $2\kpc$, $M_{\rm DM}(<2\kpc)$. Now from our dynamical modelling, we can adapt the value of $M_{\rm DM}(<2\kpc)$ to what would be required in order to best match the \brava velocity dispersions. This is done iteratively during the \nmagic fit: every eight time units we evaluate the value of a factor denoted by $\mathcal{F}$ to be applied on the model velocity dispersions in the \brava fields in order to best fit the data. Heuristically, $\mathcal{F}^2$ estimates the multiplicative factor to be applied to the total dynamical mass in the bulge in order to best fit the \brava dispersions. Thus, $\mathcal{F}^2>1$ ($<1$) indicates that more (less) mass in the bulge is required to get a better agreement with the data. Therefore, every eight time units we increase $M(<2\kpc)$ by $\Delta M(<2\kpc) = 1.0 \times 10^{10} \, \Msun \times (\mathcal{F}^2-1)$ and redefine the dark matter density in the entire galaxy by fitting again the Einasto density to the rotation curve data and the new value for $M_{\rm DM}(<2\kpc)$. The pre-factor of $10^{10} \, \Msun$ sets the rate at which the inner dark matter is adapted, such that at end of the fit $M(<2\kpc)$ has converged to the \brava dispersion for the considered pattern speed and stellar mass distribution. We assume that the halo is an oblate spheroid with a vertical flattening of $0.8$ throughout the Galaxy. There is not enough power in our data to see a significant effect of the dark matter shape that is not degenerate with the dark matter profile and therefore more complicated 3D shapes of the dark matter density are not justified. In practice, we expand the target halo density in spherical harmonics up to order eight and fit the dark matter particles to the expansion with the M2M method using the spherical harmonics kernels extensively described in previous uses of \nmagic (see \citealt{DeLorenzi2007} and subsequent work for more detail).

\subsection{The missing central mass}
\label{section:missingCentralMass}

\begin{figure}
  \centering
  \includegraphics{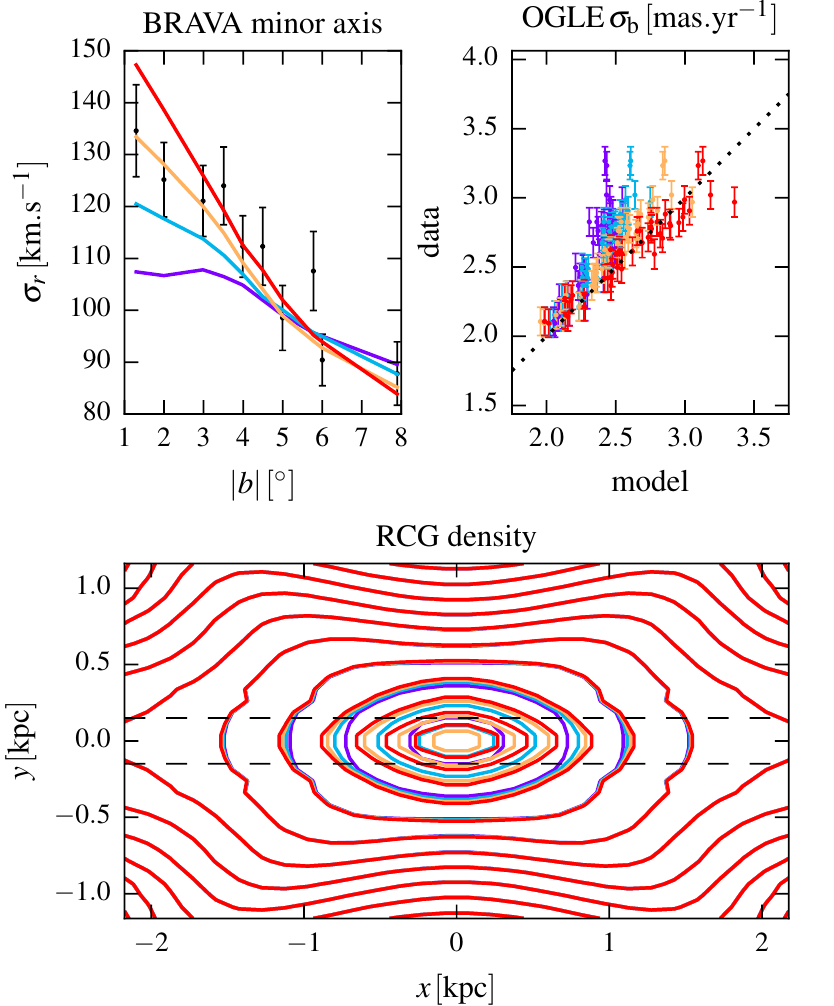}\\
  \caption{Illustration of the effect of the extra central mass on the \brava dispersions along the minor axis (upper left) and on the proper motion dispersion in the $b$ direction (upper right) for four models with no extra central mass (purple) or an extra central bar mass of $0.1 \times 10^{10} \, \Msun$ (blue),  $0.2 \times 10^{10} \, \Msun$ (yellow) and $0.3 \times 10^{10} \, \Msun$ (red). The lower plot shows the side-on projection of the peanut shape assuming that the extra mass is stellar. Most of the extra mass is located in the missing $\pm150\pc$ strip and therefore does not violate the measurement of the peanut shape from RCGs.}
  \label{fig:effectCentralMass}
\end{figure}

Using the modelling procedure defined above and the best dark matter halo that matches the overall \brava dispersion, we find evidence for a lack of central mass concentration in the models. In \autoref{fig:effectCentralMass} the purple line and points show the model predictions of the \brava dispersions along the minor axis and the \ogle proper motion dispersions along the $b$ direction for our fiducial bulge density and a pattern speed of $40\kmskpc$. The underestimation of the central dispersion is very clear, and very little freedom is available either in the mass-to-clump ratio or in the pattern speed to correct for this. As \nmagic adjusts the orbital structure in order to best match the kinematics, the only way remaining to increase the central dispersion is to deepen the gravitational potential by adding an extra central mass component. Motivated by the massive nuclear disk found by \citet{Launhardt2002} and considering the fact that the vertical structure of the bulge density could be steeper than our fiducial $\rm{sech}^2$ extrapolation, we model the missing mass by assuming that it is distributed as an elongated exponential disk following the bar, whose density is given by
\begin{equation}
 \rho_{\rm c} \propto \exp \left ( -\frac{\sqrt{x^2 + (y/0.5)^2}}{h_{\rm r}} \right) \times \exp \left ( -\frac{|z|}{h_{\rm z}}\right )
 \label{equation:centraldensity}
\end{equation}
where $x$, $y$, and $z$ are coordinates along the principal axes of the bar in $\kpc$. When $h_{\rm r}$ and $h_{\rm z}$ are too small, the large central concentration leads to the death of the peanut shape as already noted by \citet{Athanassoula2005a}. When $h_{\rm r}$ and $h_{\rm z}$ are too large, the central concentration is not sufficient to provide the central potential we need to match our kinematic data. After experimenting with several combinations of $h_{\rm r}$ and $h_{\rm z}$, we adopt the values of $250\pc$ and $50\pc$, respectively. 

The other three lines in \autoref{fig:effectCentralMass} show the predictions of models containing an extra bar-like component with the density of \autoref{equation:centraldensity} normalized to masses of $M_c = 0.1$, $0.2$ and $0.3\times 10^{10} \, \Msun$. We see that we need an additional mass of about $0.2\times 10^{10} \, \Msun$ in order to reproduce both the \brava minor axis dispersions and the \ogle proper motions. In the lower plot of \autoref{fig:effectCentralMass} we show the side-on projection of the RCG density in the bulge assuming that this extra mass is stellar and traced by red clump stars. With our parametrization, most of the extra central component is located in the inner mid-plane $\pm 150\pc$ strip where the RCG density has not been directly measured. The interpretation of this extra mass is discussed in more detail in \autoref{section:discussionCentralMass}.

\section{Key parameters of the Milky Way's effective potential}
\label{section:Results}

In this section we explore systematically the key parameters of the Milky Way's effective potential. We focus on the three parameters that have the largest impact on our data sets: (i) the bar pattern speed $\Omega_b$, (ii) the mass-to-clump ratio $\rm{M/n_{RCG}}$ and (iii) the extra central mass noted $M_{\rm c}$. We explore nine values of $\Omega_b$ between $25\kmskpc$ and $50\kmskpc$ in steps of $2.5\kmskpc$, three values for $\rm{M/n_{RCG}}$, $1000$ (our fiducial) and $900$ and $1100$ corresponding to the range of values consistent with the bulge stellar population (see \autoref{section:empiricalMassToClump}), and five values for $M_c$ between $0.1\times 10^{10}\, \Msun$ and $0.3\times 10^{10}\, \Msun$ in steps of $0.05\times 10^{10} \, \Msun$. This results in a 3D grid of 135 \nmagic runs for which we will evaluate how well they reproduce each individual data set.

\begin{figure}
  \centering
  \includegraphics{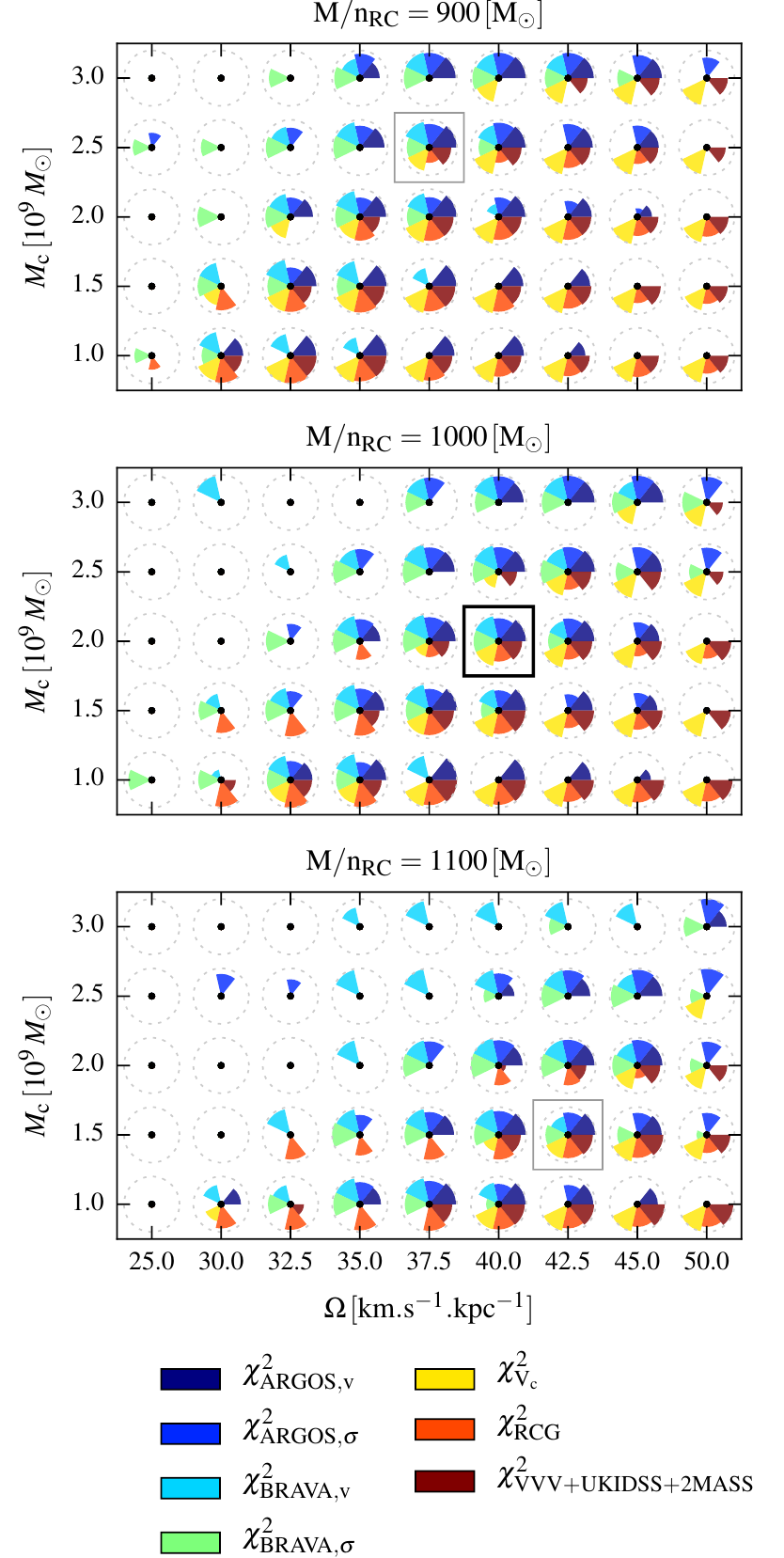}\\
  \caption{The cheese plot: overview of the systematic search of the 3D parameter space $(\rm{M/n_{RCG}}, M_c, \Omega_b)$. Each panel shows a slice in the 3D parameter space for a mass-to clump ratio of 900 (top), 1000 (middle) and 1100 (bottom). Each black dot corresponds to one simulation and the area of each color wedge is a measure of how well a simulation reproduces the different data sets. We identify our best model at $(\rm{M/n_{RCG}}, M_c, \Omega_b) = (1000, 0.2 \times 10^{10} \, \Msun, 40\kmskpc)$ as shown by the solid black square. The two models in grey squares are our two boundary models identified in \autoref{section:ResultsCentralMass}.}
  \label{fig:cheesePlot}
\end{figure}

In \autoref{fig:cheesePlot} we give an overview of our 135 simulations. In this figure, the area of the different wedges for each model shows how well that simulation performs in reproducing the various data set, larger area meaning better agreement. The area of each wedge is proportional to $1-\tilde{\chi}^2$ where $\tilde{\chi}^2$ is the $\chi^2$ rescaled so that the best simulation has $\tilde{\chi}^2$ of 0 and the median simulation has $\tilde{\chi}^2$ of 1, separately for each observable. Some interesting features are directly visible in \autoref{fig:cheesePlot}. From the \argos and \brava kinematics, we see that the mass-to-clump ratio has some degeneracy with both the pattern speed and the central mass: low mass-to-clump ratio tends to prefer lower pattern speed and higher central mass. The model at $(\rm{M/n_{RCG}}, M_c, \Omega_b) = (1000, 0.2 \times 10^{10} \, \Msun, 40\kmskpc)$ as indicated by the black square in \autoref{fig:cheesePlot} is able to reproduce well all data sets simultaneously and therefore constitutes our best model.

In order to recover the range of $\rm{M/n_{RCG}}$, $M_c$ and $\Omega_b$ around the best model that is allowed by the data, we would ideally construct a global likelihood for all our data sets together and search for the 3D region of the grid that contains the $68\%$ most likely models. However, this approach is not applicable here for two reasons:
\begin{enumerate}
  \item Comparing different data sets in a purely statistical way is dubious when the fitted data sets, such as the RCG density in the bulge and the \ogle proper motions, have dominant errors that are not statistical. We are more interested in reproducing all data sets fairly well at once than in maximizing the formal total likelihood, which may be dominated by one particular aspect because of greatly different number of observables or unaccounted systematic errors.

  \item The evaluation of the range of models consistent with some data within some confidence limit is not a straightforward task in M2M modelling. \citet{Morganti2013} showed that the magnitude of the $\Delta \chi^2$ that allows a fair estimation of the range of models that is compatible with the data can be much larger than that expected, due to modelling noise.
\end{enumerate}

Instead, we adopt a more phenomenological approach where we use our knowledge of the bar dynamics gained from the experiments of \autoref{section:bulgeDynamics} to explore the 3D grid of models in more detail. In \autoref{section:ResultsBarPatternSpeed}, we study the mass-to-clump/pattern speed degeneracy and recover the pattern speed of the galactic bar. In \autoref{section:ResultsCentralMass}, we study the mass-to-clump/central mass degeneracy and recover the best value for the central mass. We finally show how our best model compares to the different data sets in \autoref{section:BestModel}.

\subsection{Bar pattern speed and corotation}
\label{section:ResultsBarPatternSpeed}

\begin{figure}
  \centering
  \includegraphics{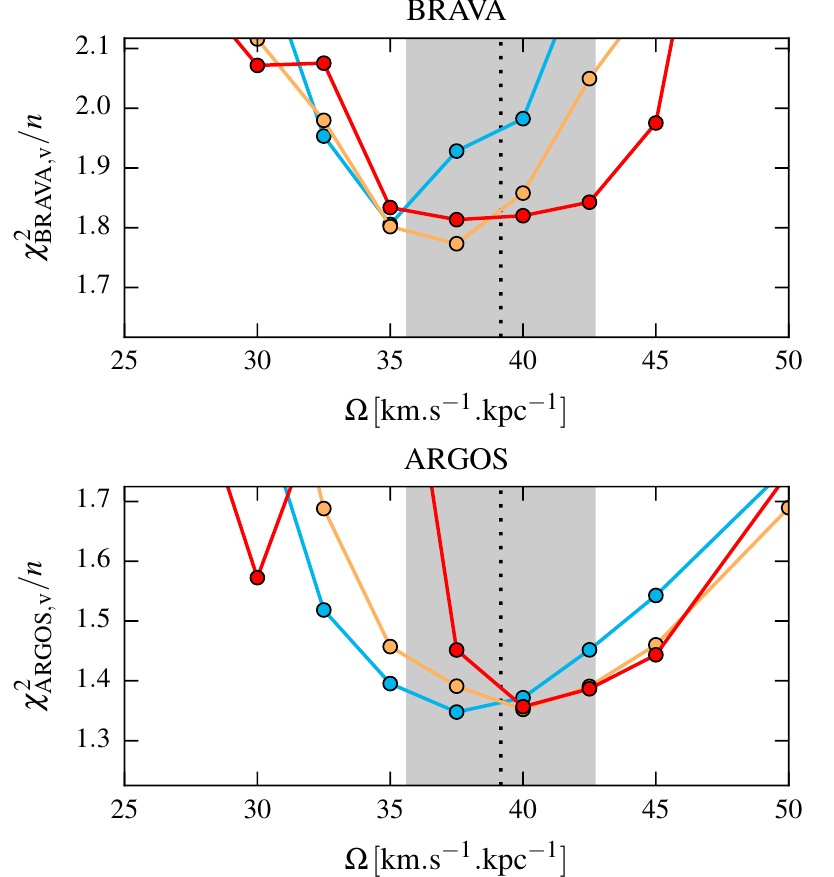}\\
  \caption{$\chi^2$ per datapoint as a function of the pattern speed $\Omega_b$ for the \brava (top) and \argos (bottom) mean velocities. The blue, yellow and red lines correspond respectively to mass-to-clump ratio of $\rm{M/n_{RCG}}=900$, 1000 and 1100, all for their respective best value for the central mass $M_c$ whose determination is described in \autoref{section:ResultsCentralMass}. The grey area spans the one sigma error range of our pattern speed estimation of $39\pm3.5\kmskpc$.}
  \label{fig:chi2patternSpeed}
\end{figure}

We showed in \autoref{section:patternSpeedAndDarkHalo} that the pattern speed had a clear signature in the mean radial velocity observables. Hence, we focus on the mean velocity of both the \brava and \argos surveys. For each combination of pattern speed and mass-to-clump ratio, we first search for the best value of $M_c$, as described in the next subsection. In \autoref{fig:chi2patternSpeed}, we show the $\chi^2$ per datapoint as a function of the pattern speed $\Omega_b$ for different mass-to-clump ratios and the corresponding best value of $M_c$. Good fits to the data are found in the range $\Omega_b \sim 35-42.5\kmskpc$ depending on the mass-to-clump ratio. An increase in mass-to-clump ratio of $10\%$ requires an increase in $\Omega_b$ of $\sim2.5\kmskpc$. We also notice that the \argos data systematically prefers pattern speeds $\sim2.5\kmskpc$ larger than the \brava data. Forming a joined $\chi^2$ between \argos and \brava and assuming flat priors on the mass-to-clump ratio in the range $900-1100$, we find a mean of the best pattern speeds to be $\Omega_b = 39\kmskpc$. As already discussed, the evaluation of statistical errors on measurements and parameters from M2M modelling is usually problematic since the classical $\Delta \chi^2 = 1$ method tends to underestimate the real error. \citet{Morganti2013} developed a method to better estimate the statistical error using a value of $\Delta \chi^2$ corresponding to the scatter of the $\chi^2$ surface from the models around the minimum. Using this method we find statistical errors lower than $1 \kmskpc$, smaller than the systematics arising from both the degeneracy with the mass-to-clump ratio and the systematic offset between the \argos and \brava data sets. Adding in quadrature an error of $2.5\kmskpc$ from both these sources of systematics, we conclude that the pattern speed of the Milky Way bar is $\Omega_b = 39\pm 3.5\kmskpc$. Using the composite rotation curve of \citet{Sofue2009} rescaled to $(R_0, V_0) = (8.2\kpc, 238\kms)$, the corotation radius of the bar is found at $R_{\rm{cr}} = 6.1\pm0.5\kpc$.

\subsection{Central mass distribution}
\label{section:ResultsCentralMass}

\begin{figure}
  \centering
  \includegraphics{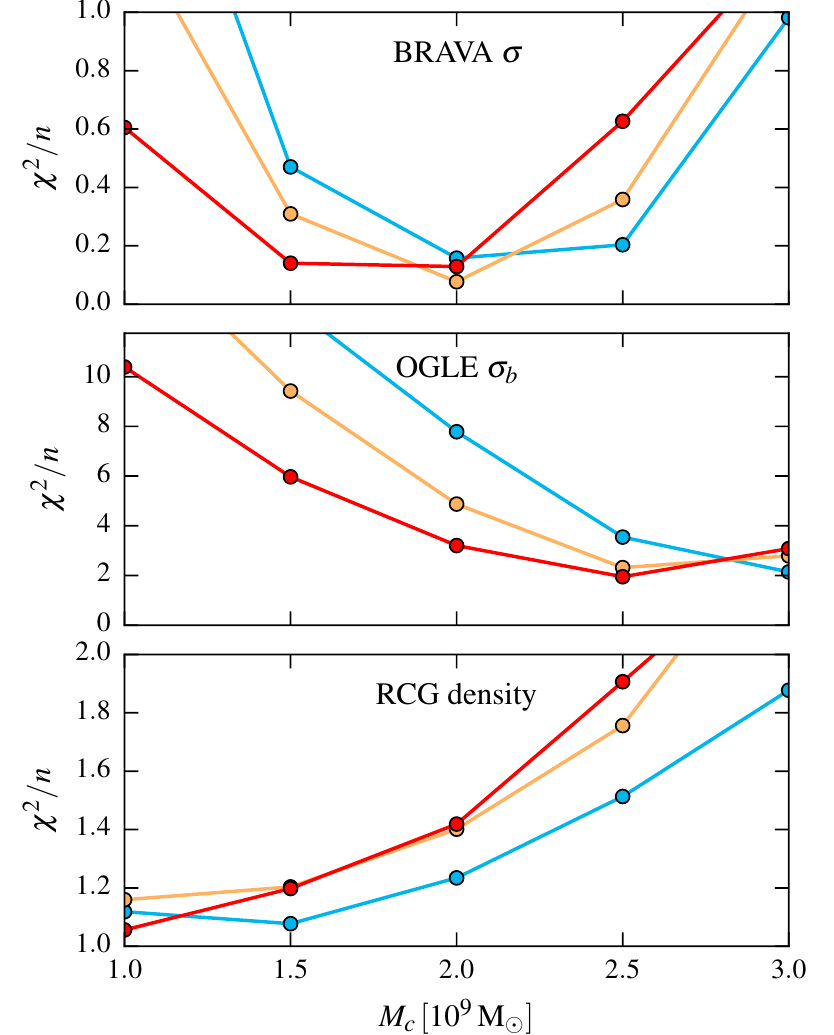}\\
  \caption{$\chi^2$ per datapoint as a function of the additional central mass $M_c$ from the central \brava dispersions (top) and the \ogle proper motions in the $b$ direction (middle) and the RCG density (bottom). The blue, yellow and red lines correspond respectively to mass-to-clump ratio of $\rm{M/n_{RCG}}=900$, 1000 and 1100, all for their respective best value for the pattern speed determined as in \autoref{section:ResultsBarPatternSpeed} from a joined $\chi^2$ between \argos and \brava. Proper motion tends to prefer large central masses but larger masses tend to weaken or destroy the peanut shape of the bulge.}
  \label{fig:chi2Mc}
\end{figure}

We showed in \autoref{section:missingCentralMass} the necessity of an additional central mass component $M_c$ for matching the inner \brava dispersions and \ogle proper motions in the $b$ direction. In \autoref{fig:chi2Mc}, we show the $\chi^2$ per datapoint of the best pattern speed models as a function of $M_c$ for all the \brava and \ogle fields within $3.5\degree$ from the centre ($\sim 500\pc$ at the distance of the GC). We see that the \brava radial velocity prefers an additional mass of $0.2 \times 10^{10} \, \Msun$ with a slight degeneracy with the mass-to-clump ratio where a $10\%$ increase in $\rm{M/n_{RCG}}$ leads to a decrease of $M_c$ of about $0.05 \times 10^{10} \, \Msun$. The signature of $M_c$ in the $\sigma_b$ proper motions appears to be in slight tension with the \brava dispersion, having a systematic preference for higher values of $M_c$. Note however that these proper motions are simply predictions and thus are not expected to perfectly fit the data. We show in \autoref{fig:bestModels_ogle} that for a central mass of $0.2 \times 10^{10} \, \Msun$ our best model provides a very good fit to the proper motion dispersions in both directions, staying on average within 5\% of the data, even though it systematically underestimates it. In addition, by looking at the $\chi^2$ of the RCG density as shown in the bottom plot of \autoref{fig:chi2Mc}, we find that large values of $M_c$ tend to weaken or destroy the peanut shape of the bulge, as was already shown by previous studies (e.g \citealt{Shen2004, Athanassoula2005a}).
Since no value of $M_c$ simultaneously gives a best fit to the \ogle $\sigma_b$ predictions and the RCG density fit, we estimate $M_c$ based on the \brava central dispersions to be $M_c = 0.20  - 0.05 \times (\rm{M/n_{RCG}}-1000)/100  \, \times 10^{10} \, \Msun$ and recognize the presence of possible unaccounted systematic effects.

Since a $10\%$ increase in mass-to-clump ratio requires a $2.5\kmskpc$ increase in pattern speed and a $0.05 \times 10^{10} \, \Msun$ decrease in central mass we define two boundary models around our best model at each end of this three dimensional degeneracy, $(\rm{M/n_{RCG}}, M_c, \Omega_b) = (900, 0.25 \times 10^{10} \, \Msun, 37.5\kmskpc)$ and $(\rm{M/n_{RCG}}, M_c, \Omega_b) = (1100, 0.15 \times 10^{10} \, \Msun, 42.5\kmskpc)$. Those two models are indicated by the grey squares in \autoref{fig:cheesePlot} and are used in \autoref{section:stellarAndDarkMatterMass} to quantify uncertainties in measuring the stellar and dark matter mass distributions.

\subsection{Best fitting dynamical model of the Galaxy}
\label{section:BestModel}

\begin{figure*}
  \centering
  \includegraphics{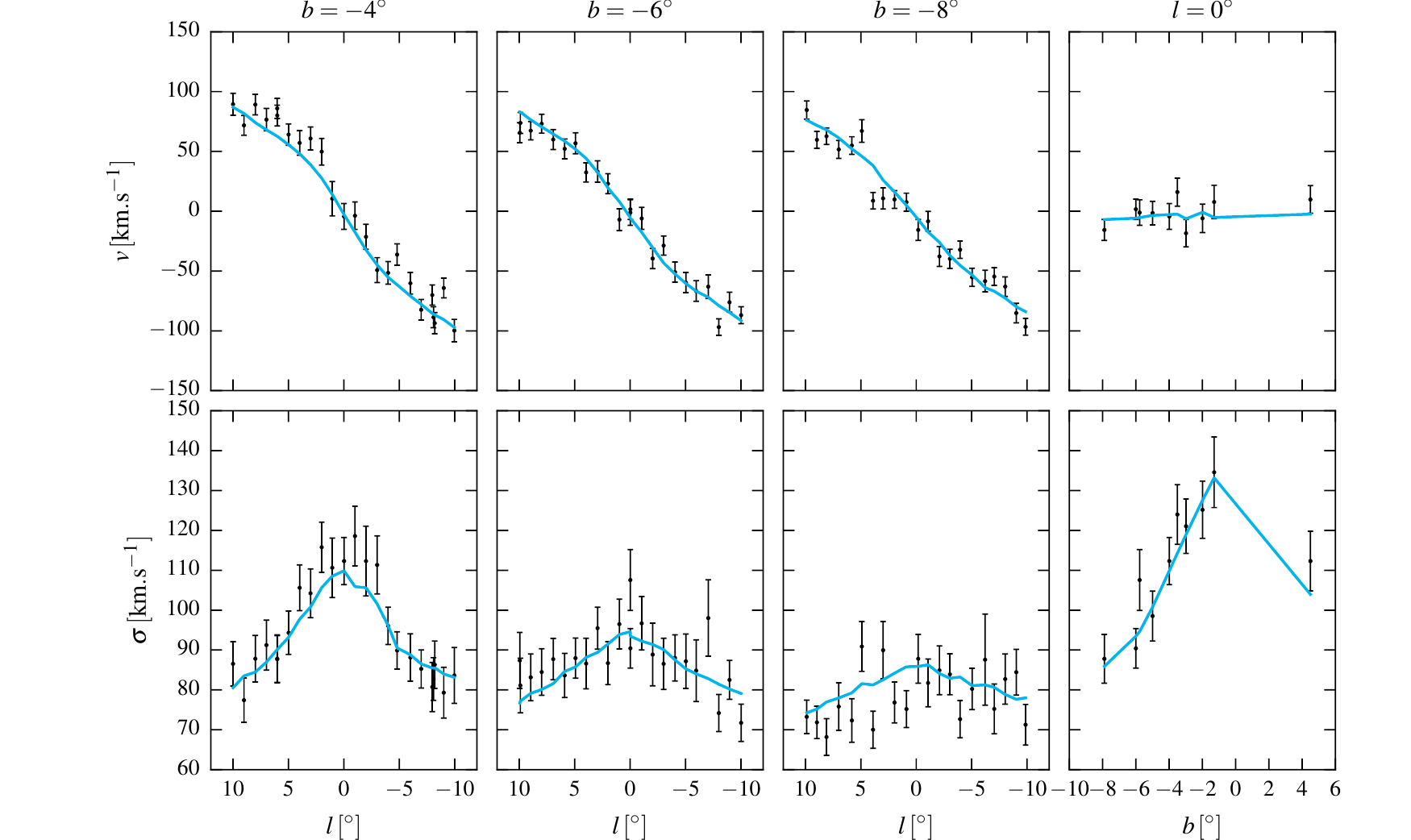}\\
  \caption{Comparison of our best model with the \brava mean radial velocities (upper row) and velocity dispersions (lower row) for latitude slices at $-4\degree$, $-6\degree$ and $-8\degree$ and along the minor axis.}
  \label{fig:bestModels_brava}
\end{figure*}
Starting with the bulge kinematics, \autoref{fig:bestModels_brava} shows how our best model compares to the \brava mean velocities and velocity dispersions. The agreement is overall very good and is an improvement over \hyperlink{P15}{P15} where the dispersions at $-6\degree$ were always underestimated. 

\begin{figure}
  \centering
  \includegraphics{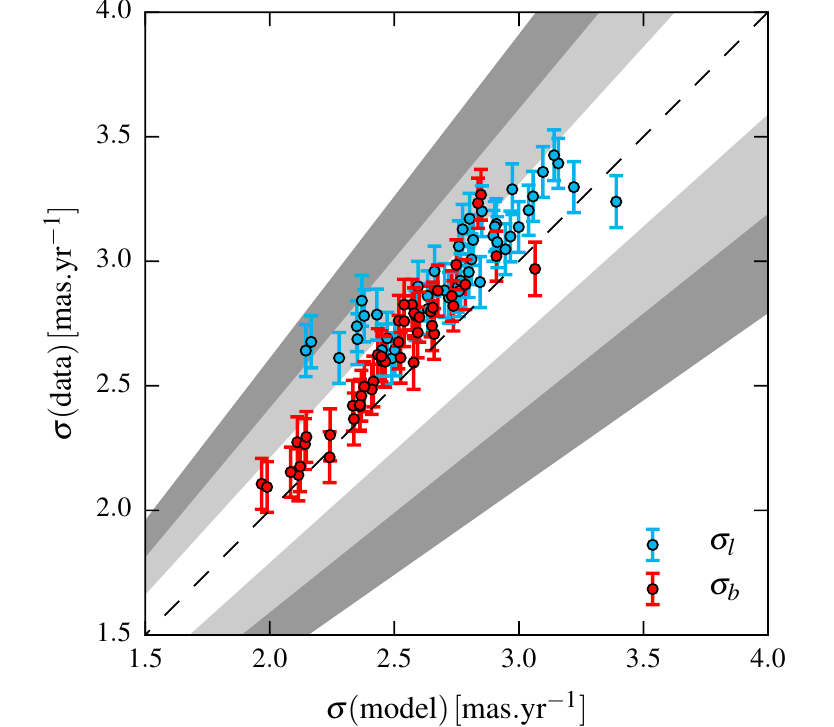}\\
  \caption{Model/data comparison of the \ogle proper motion dispersion in the $l$ (blue) and $b$ (red) directions. Shaded areas indicate fractional differences between 10 and 20\%, and between 20 and 30\%. Nearly all our best model proper motions are within 10\% of the data.}
  \label{fig:bestModels_ogle}
\end{figure}
The largest improvement over \hyperlink{P15}{P15} is seen in the proper motion dispersions in the $l$ and $b$ directions shown in \autoref{fig:bestModels_ogle}. In nearly all fields, the model is within 10\% of the \ogle proper motions even though the proper motion data were not included in the fitting procedure. We notice however that the model tends to systematically underestimate the data by about $5\%$. An increase in $\sigma_b$ can be obtained by increasing the central mass but at the cost of losing the peanut shape; or by lowering the pattern speed at the cost of a worse agreement with the mean radial velocity. In the $l$ direction, an increase in $\sigma_l$ can be obtained by increasing the bar angle, at the cost of a worse fit to the RCGs histograms in the bar. Since a $10\%$ variation, i.e. $0.2-0.3\masyr$ is only slightly larger than the field-to-field variation of the data ($\sim 0.2\masyr$, \citealt{Rattenbury2007}), we consider our model as already consistent with the \ogle proper motions.  

\begin{figure}
  \centering
  \includegraphics{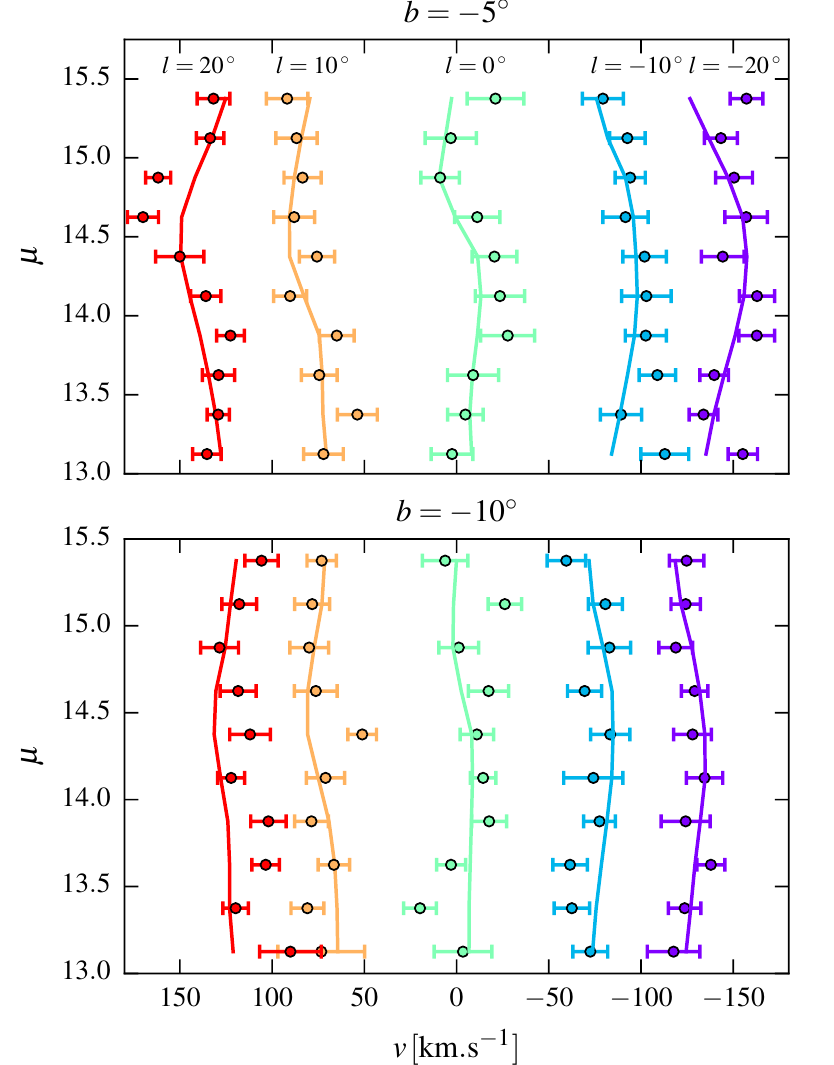}\\
  \caption{\argos mean radial velocities as a function of distance modulus compared with our best model, at $b=-5\degree$ (upper panel) and $b=-10\degree$ (lower panel). Different colors show different longitudes spaced every $10\degree$ from $-20\degree$ (purple) to $+20\degree$ (red). The curvature along the minor axis field (green) is a clear signature of the streaming motions in the bar.}
  \label{fig:bestModels_argos}
\end{figure}

\begin{figure*}
  \centering
  \includegraphics{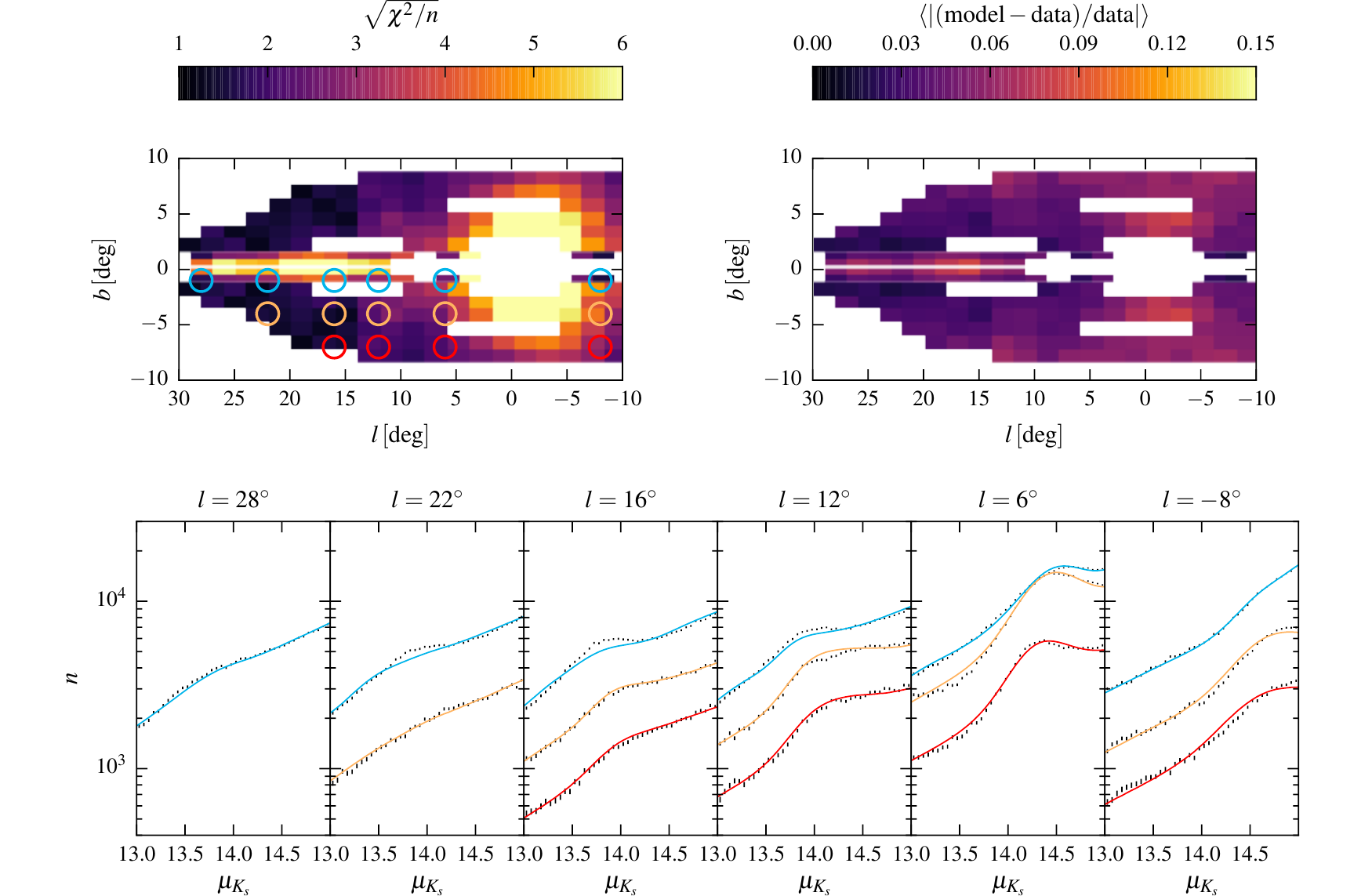}\\
  \caption{Red clump magnitude distribution of our best model across the bulge and long bar region. The upper-left panel shows the reduced $\chi^2$ between the model and the data along all considered lines of sight and the upper-right panel the mean fractional errors. The $\chi^2$ is formally poor in the bulge and close to the plane due to the very low statistical errors arising from the large number of observed stars. The lower six panels show the model line of sight magnitude distributions on top of the data for six different latitudes at three heights below the plane as indicated by the colored circles in the top-left panel.}
  \label{fig:bestModels_RCHistograms}
\end{figure*}

Extending to the bar region, \autoref{fig:bestModels_argos} shows the \argos mean velocity as a function of distance. The streaming motion along the bar is very visible at latitude $b=-5\degree$ from the twist in the mean velocity as a function of distance. Here again the model performs very well in reproducing the data.
The transition between the bulge and the bar, together with the in-plane structure of the bar, is shown in \autoref{fig:bestModels_RCHistograms}.
The model does a very good job at reproducing the RCG distribution for $|b|\geq2\degree$. In the plane, and mostly in the region $12\degree\leq l\leq22\degree$, the model cannot find a good fit to the very narrow distribution of RCGs along the line of sight. This is probably due to the superthin bar component found by \hyperlink{W15}{W15}. The investigation of the detailed structure of the superthin bar is beyond the scope of this paper but could be addressed in the future using the \apogee data. The \apogee survey, as an infrared spectroscopic survey, can penetrate the high extinction in the plane and provides stellar kinematics and chemical abundances in the superthin bar region. By looking at clumps in chemical space, \citet{Hogg2016} already found evidence for a very young stellar component that may correspond to the superthin bar of \hyperlink{W15}{W15}. 
\begin{figure*}
  \centering
  \includegraphics{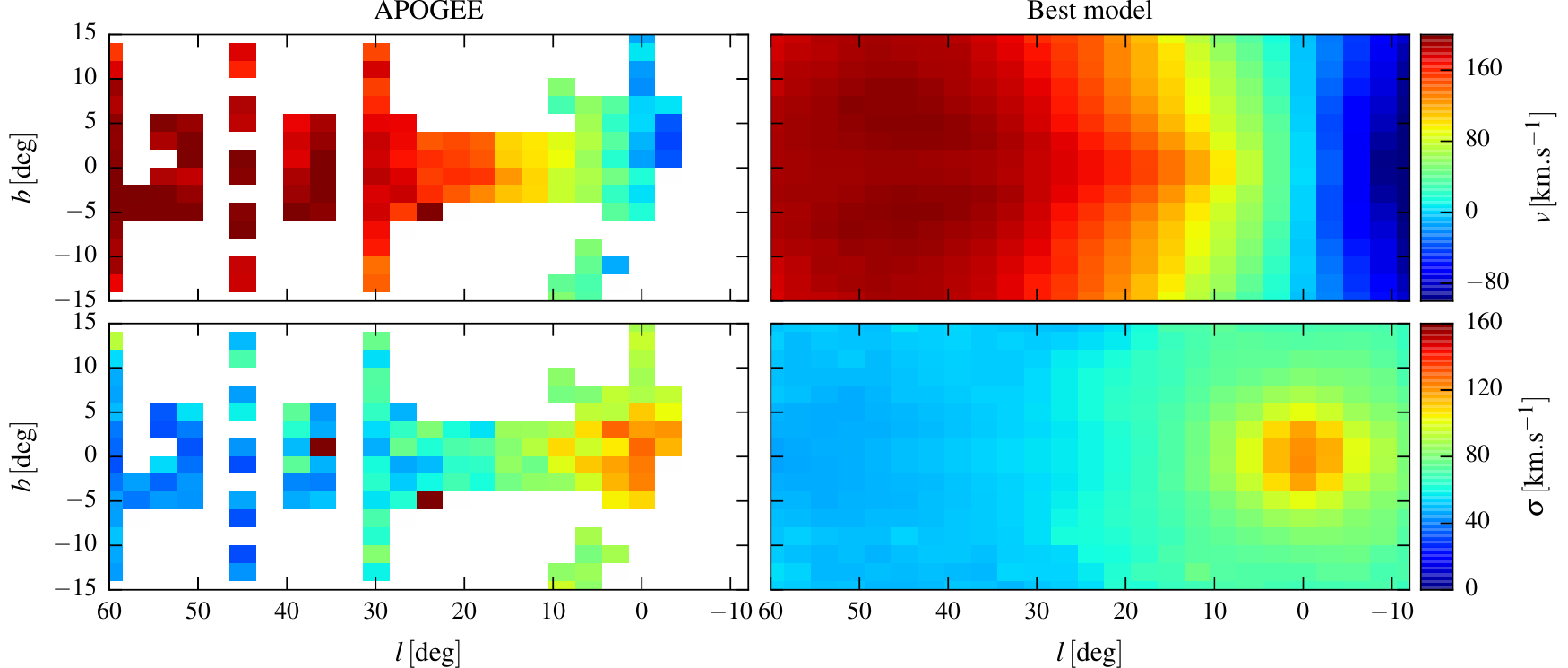}\\
  \caption{Comparison of the mean velocity (top) and velocity dispersion (bottom) between the latest \apogee kinematics from \citet[][left]{Ness2016} and our best model (right). Velocities are expressed in the galactic inertial frame and only stars between $4$ and $12\kpc$ along the line of sight are taken into account. Even though not fitted outside the bar region, the model is already in good agreement with the \apogee kinematics.}
  \label{fig:bestModels_Apogee}
\end{figure*}
Anticipating the future use of \apogee to constrain the dynamical models further, we show in \autoref{fig:bestModels_Apogee} a comparison between our best model and the latest \apogee kinematics from \citet{Ness2016}. Our model is already in very good qualitative agreement with \apogee even though a more quantitative comparison would require the modelling of the \apogee selection function which is not included here.

\section{Stellar and dark matter mass distribution in the Milky Way}
\label{section:stellarAndDarkMatterMass}

In this section we study the stellar and dark matter mass distribution of our best model and evaluate the effect of variations in the different modelling assumptions. As a range of reasonable variations around our best model, we will consider the following:
\begin{itemize}
 \item The two boundary models of \autoref{section:Results} found at each end of the three dimensional degeneracy valley between mass-to-clump ratio, bar pattern speed and additional central mass;
 \item Varying the bar angle from our fiducial $28\degree$ to either $23\degree$ or $33\degree$;
 \item Varying the dark matter flattening from our fiducial $0.8$ to either $0.6$ or $0.4$;
 \item Varying the outer disk scalelength from our fiducial $2.4\kpc$ to either $2.15\kpc$ \citep{Bovy2013} or $2.6\kpc$ \citep{Juric2008};
\end{itemize}
These variations will be used to evaluate errors (systematic) on the mass parameters found from our best model. All errors quoted in the next two sections refer to half the range of values found in the fiducial and variation models. A summary of the mass parameters of our best model is given in \autoref{table:finalMassParameters}.

\subsection{Bulge, bar, inner and outer disk}
\label{section:ResultsStarsInBar}

\begin{figure}
  \centering
  \includegraphics{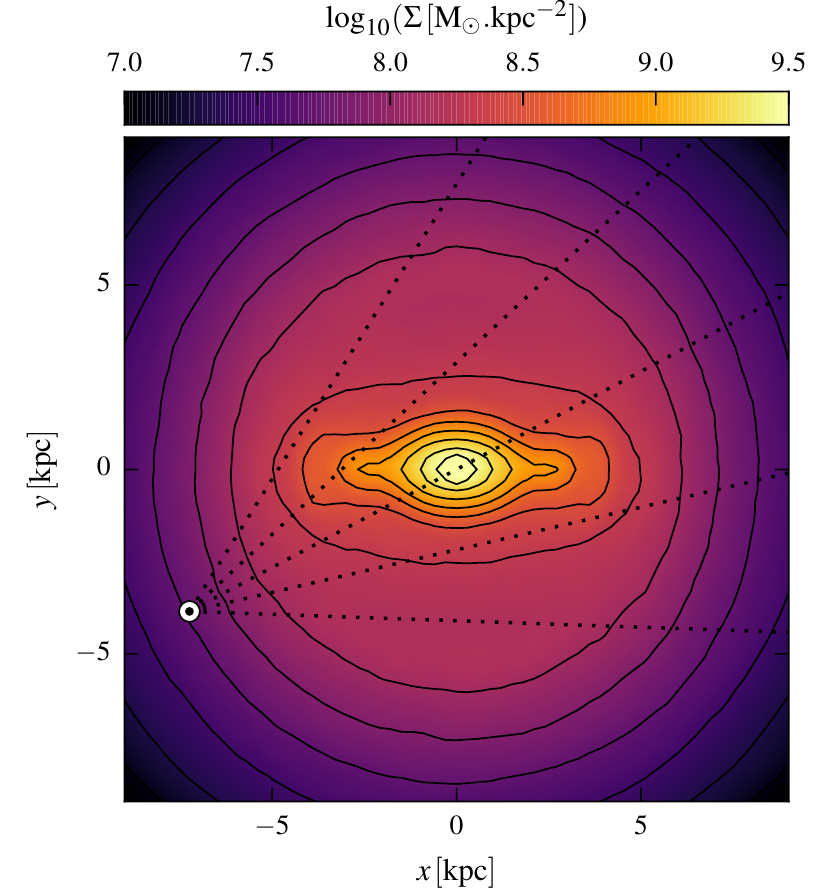}\\
  \caption{Surface density map of our best model. The bar extends to $5.3\kpc$ from the centre and rotates at $\Omega_b = 40\kmskpc$. The dotted lines originating from the Sun (dot symbol) indicate sight lines with galactic longitudes $l=-30\degree, -15\degree, 0\degree, +15\degree$ and $+30\degree$.}
  \label{fig:SurfaceDensityBestModel}
\end{figure}

\begin{figure}
  \centering
  \includegraphics{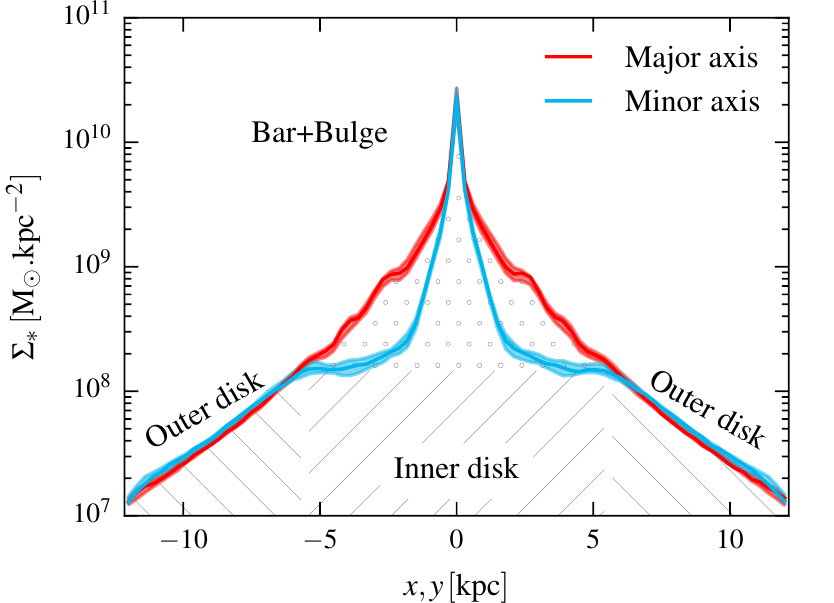}\\
  \caption{Stellar surface density profiles along the bar major axis (red) and bar in-plane minor axis (blue) for our best model (solid line) and the variation models (color span). The hatched region indicates the bar+bulge component (circle) the inner disk (/) and the outer disk (\textbackslash) according to our photometric definition of the components.}
  \label{fig:SurfaceDensityProfile}
\end{figure}

In \autoref{fig:SurfaceDensityBestModel} and \autoref{fig:SurfaceDensityProfile}, we show the surface density map and profiles of our best model with the range of the model variations, assuming that the additional mass component is stellar as discussed in \autoref{section:discussionCentralMass}. The entire Galaxy is to some level non-axisymmetric. This is very clear at the edge of the bulge where the surface density along the minor axis at $2\kpc$ is a factor of about 4 smaller than along the major axis. Both profiles cross at $6.3\kpc$ from the centre, close to the corotation radius. Beyond corotation, the surface density becomes larger along the minor axis than along the major axis, as one would expect based on the linear perturbation of near circular orbit \citep{Dehnen2000a, Binney2008}. Along the major axis, the density of the bar is close to exponential. This is a characteristic of late-type barred galaxies as revealed by the $\rm{S^4G}$ survey \citep{Elmegreen2011}. Traditionally, the complex structure of the stellar density of the Galaxy has been divided into a small number of discrete components, mainly bulge, bar and disk for which we would wish to measure mass and shape. We will focus here on three definitions:
\begin{enumerate}
 \item Bar, bulge and disk structure from stellar mass `photometric' profiles: From the major and minor axis profiles in \autoref{fig:SurfaceDensityProfile}, we see clearly three regimes: (a) the outer disk, nearly axisymmetric and exponential outside $\sim5.3\kpc$ (b) the inner disk, axisymmetric with a nearly constant surface density inside $5.3\kpc$ (c) the bar/bulge, i.e. the bar which formed a bulge in its inner part. This photometric definition of the bar and inner disk components has the advantage to be easily applicable to external galaxies. By integrating the surface density associated with the `photometric' bulge and bar, we find a `photometric' bar/bulge stellar mass of $M_{\rm{bar/bulge}} = 1.88 \pm 0.12 \times 10^{10}\, \Msun$, among which $1.34 \pm 0.04 \times 10^{10}\, \Msun$ is located in the bulge and $0.54 \pm 0.04 \times 10^{10}\, \Msun$ in the long bar.  The `photometric' stellar mass associated with the inner disk within the bar region is found to be $M_{\rm{inner\ disk}} = 1.29\pm0.12\times10^{10}\, \Msun$, lower than the mass of the bar/bulge structure. The errors in these quantities are systematics derived from the variation models presented previously.
 
 \item Non-axisymmetric long bar from 2D `photometry': An alternative way to define the bar component is to search for non axisymmetries in the face-on surface density. We first define a maximum axisymmetric model using the minor axis profile and remove this maximum axisymmetric model from the original surface density map. By integrating the residuals for radii inside $5.3\kpc$ and outside the bulge, we find a non-axisymmetric long bar mass of $M_{\rm{non-axi}} = 0.46 \pm 0.03 \times 10^{10}\, \Msun$. This measure is similar to but slightly lower than the previous `photometric' long bar mass since the inner disk has a similar but slightly lower surface density than our maximum axisymmetric model. From both estimates combined, the bar outside the bulge has slightly less than half the mass of the disk in the same radial range.
 
 \item Bar-following orbits: Unlike some spiral structures, the bar is not a density wave: stars that form the bar stay in the bar and orbit mostly along elongated orbits of the $x_1$ family and its descendants \citep{Contopoulos1989}. Since we have access to individual orbits in our dynamical model, we can directly identify orbits that compose the bar using the method of \citet{Portail2015b}. We integrate all particles and compute the dominant frequencies of the time variation of the cylindrical radius $f_r$ and bar major axis position $f_x$ in the corotating frame. Bar-supporting orbits are found in the vicinity of $f_r/f_x = 2$, i.e have two radial oscillations for one period along the bar. In the bar region, particles that do not follow the bar have Rosetta-like orbits in the bar frame for which $f_r/f_x \neq 2$; they build the inner disk. This orbit-based definition of the bar is more elegant than the `photometric' definition and is also closer to what makes the bar a separate component but is in general not observable in external galaxies for which individual orbits are usually unknown. We find a stellar mass on bar-supporting orbit of $1.04 \pm 0.06 \times 10^{10} \, \Msun$. This estimate misses the non-bar following orbits in the bulge.
\end{enumerate}

\hyperlink{W15}{W15} determined from a combination of \vvv, \ukidss and \twomass data the length of the bar and found a half-length of $5.0\pm0.2\kpc$. Since our model is the first non-parametric fit of the galactic bulge and bar it is important to see how it compares to direct determination from the data. We follow \hyperlink{W15}{W15} and focus on the following three methods to measure the bar half-length:
\begin{itemize}
 \item $L_{\rm{drop}}$: radius at which the ellipticity of the bar drops the fastest;
 \item $L_{\rm{prof}}$: radius at which the major and minor axes agree within 30\%;
 \item $L_{\rm{m=2}}$: radius at which the relative m=2 component of the Fourier decomposition of the surface density drops to 20\% of its maximum value.
\end{itemize}
We measure $L_{\rm{drop}} = 5.77\kpc$, $L_{\rm{prof}} = 5.12\kpc$ and $L_{\rm{m=2}} = 5.02\kpc$. By taking the mean of those three measurements and the standard deviation of the three measurements applied on all our variation models, we find a bar half-length of the galactic bar of $5.30 \pm 0.36\kpc$, in good agreement with the measurement of $5.0\pm0.2\kpc$ found by \hyperlink{W15}{W15} from the their component fit of the RCG magnitude distributions.

\subsection{The dark matter mass distribution in the Milky Way}
\label{section:ResultsDarkMatterDistribution}

In \autoref{section:automaticDMhalo}, we showed how we recognize and evaluate the need of dark matter in the bulge from the \brava kinematics once the stellar density and bar pattern speed are fixed. Since we have a $\sim10\%$ accurate measurement of the mass-to-clump ratio in the bulge, we have access to the dark matter mass distribution in the bulge. In the volume of the box in which the RCG density was measured, i.e. a box of $(\pm 2.2 \times \pm 1.4 \times \pm 1.2 )\kpc$ along the bar principal axes, the mass budget of the bulge is found to be as follows: stars as traced by RCGs account for $1.32 \pm 0.08 \times 10^{10}\, \Msun$, dark matter accounts for $0.32 \pm 0.05 \times 10^{10} \, \Msun$ and $0.2 \times 10^{10}\, \Msun$ of additional mass are required in the centre, probably stars in the nuclear disk (see \autoref{section:discussionCentralMass}). The resulting total dynamical bulge mass is $1.85\pm 0.05\times 10^{10}\Msun$, in excellent agreement with our estimation of $1.84 \pm 0.07 \times 10^{10}\, \Msun$ found in \hyperlink{P15}{P15} from modelling the bulge only. Altogether our best model has a low dark matter fraction in the bulge of only $17\%\pm2\%$. The small error quoted here represents half the range of dark matter fractions found in the set of the fiducial and variation models considered and could underestimate the true error on the determination of the dark matter fraction in the galactic bulge. It arises because variations in the total mass, stellar mass and central mass of the bulge in our model variations approximately compensate leaving the dark matter mass in the bulge nearly unchanged.

\begin{figure}
  \centering
  \includegraphics{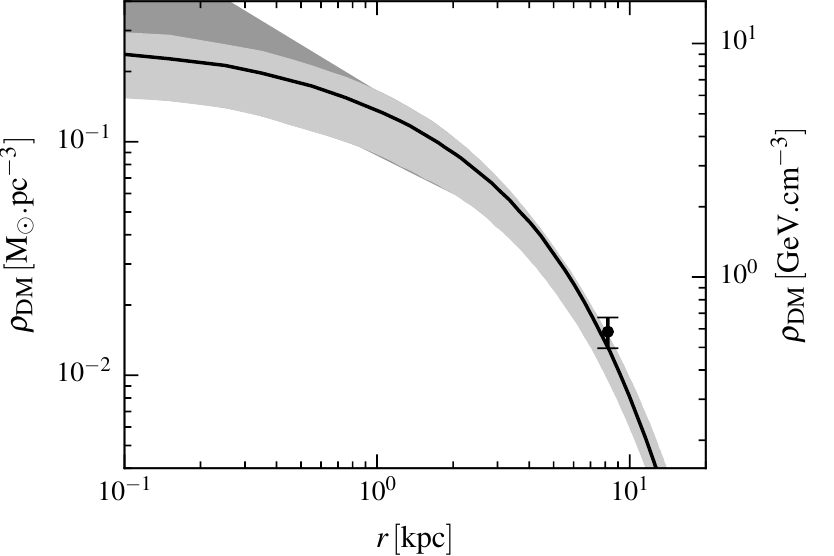}\\
  \caption{Dark matter density profile of our best model (black line), range of profiles from variation models (light grey span) and range of possible inner power-law density variations that would keep constant the dark matter mass enclosed in the bulge (dark grey span). Under the assumption of an Einasto halo, all models require a central core to account simultaneously for a low dark matter fraction in the bulge and the rotation curve at the solar radius. The datapoint at $8.2\kpc$ is the local measurement of the dark matter density from the analysis of RAVE stars from \citet{Piffl2014}, in good agreement with our best model. }
  \label{fig:bestHaloDensity}
\end{figure}

Going away from the bulge, the dark matter halo profile of our best model is shown in \autoref{fig:bestHaloDensity} together with the range of profiles from the variation models. We find a local dark matter density of $\rho_{\rm{DM}}(R_0) = 0.0132\pm0.0014\, \Msun.\pc^{-3}$, in very good agreement with the recent measurement of $\rho_{\rm{DM}}(R_0) = 0.0154\pm0.0023 \, \Msun.\pc^{-3}$ from \citet{Piffl2014} for a halo flattening of $0.8$ (see \citet{Read2014} for a review). Interestingly, our dark matter profile is cored. For the fitted Einasto density profiles, the presence of a core is a consequence of accounting simultaneously for a low dark matter mass in the bulge, a significant dark matter mass enclosed within the solar radius and a gently rising halo rotation curve. 

\begin{figure}
  \centering
  \includegraphics{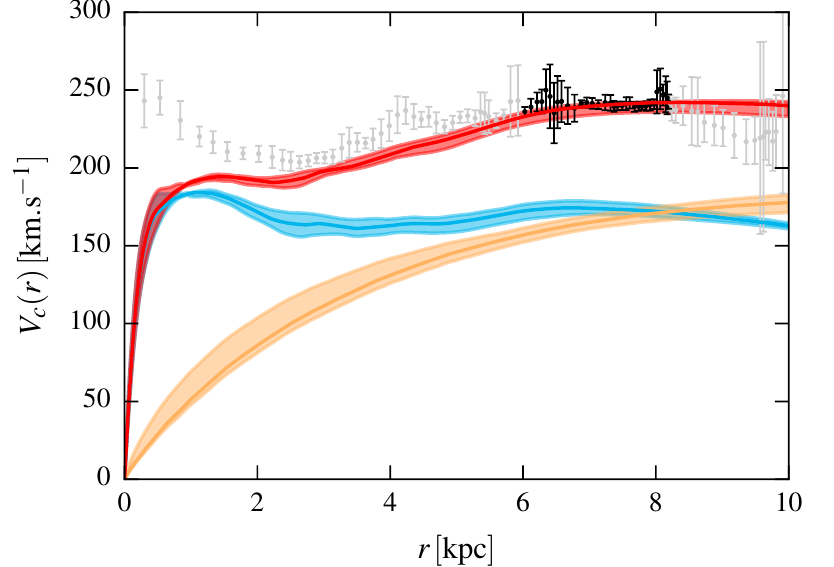}\\
  \caption{Rotation curve of our best model and range of model variations for evaluation of systematics on top of the composite rotation curve measurements from \citet{Sofue2009}. Blue, yellow and red curves represent respectively the baryonic, dark matter and total rotation curve, assuming that the totality of the additional central mass is baryonic.}
  \label{fig:bestRotationCurve}
\end{figure}

Would an NFW halo profile be allowed by our data? In the Milky Way, an NFW halo density would be expected to have a scale radius of $20-40\kpc$ \citep{BlandHawthorn2016}. Thus, in the bar region, the density would be well approximated by a simple power law $\rho_{\rm{DM}}(r)\propto r^{-1}$. Given the baryonic mass distribution of our best model and scaling the dark matter density profile such that the total circular velocity at the solar radius is matched, we find a dark matter mass in the bulge that coincidentally is in very good agreement with our best model value. However, such an NFW halo fails to reproduce the nearly flat total rotation curve observed between $6$ and $8\kpc$, with a halo circular velocity that is about $17\kms$ lower than the data at $6\kpc$ shown in \autoref{fig:bestRotationCurve}.
The presence of a core in our best model halo density thus appears as a consequence of the constraint on the flat shape of the total circular velocity in the $6-8\kpc$ range, which for our baryonic mass distribution requires the dark matter density to fall off more steeply than $\rho_{\rm{DM}}(r)\propto r^{-1}$. In order to then not overpredict the dark matter mass in the bulge, the dark halo density is forced to become shallower further in. 

For an Einsato profile, this results in a central core. However, we note that our constraint on the dark matter density in the bulge arises from a constraint on the dynamical mass of the bulge. Thus, we would also expect good agreement with the data for a steeper halo density profile in the bulge, provided it has the same dark matter mass in the bulge. In order to evaluate the possibility of a steeper density in the inner halo, we perform the following experiment. Assuming for the dark matter inside $2\kpc$ a power law density $\rho \propto r^{-\alpha}$, we determine for each model the power-law index $\alpha$ that would keep constant the dark matter mass enclosed within $2\kpc$, while ensuring continuity of the density profile at $2\kpc$. Doing so we find from our models power law slopes of $\sim -0.6$, whose range is shown as the dark grey span in \autoref{fig:bestHaloDensity}. We thus conclude that a central power-law cusp shallower than $-\alpha \sim  -0.6$ would also provide a good fit to the data.

A different view at the dark matter contribution to the gravitational potential can be seen in \autoref{fig:bestRotationCurve} where we plot the resulting rotation curve of our best model and the range of rotation curves provided by the model variations, again considering the additional central mass as entirely baryonic. The dark matter support to the rotation is often expressed as the degree of maximality of the disk \citep{Sackett1997}, representing the ratio of the stellar rotation velocity to the total rotation velocity at some particular radius. In the case of an axisymmetric disk galaxy, this radius is traditionally chosen to be $2.2$ disk scalelength, corresponding to the position of the peak of the disk rotation curve. Since the Milky Way is not axisymmetric and hosts a central bulge, the stellar rotation curve does not peak at $2.2$ outer disk scalelengths but instead at only about $1\kpc$ where a stellar rotation of $185\kms$ provides 94\% of the rotational support. At $2.2$ outer disk scalelengths, the stellar contribution to the rotation curve drops to $75\%$ still within the range of what would be called a maximum disk \citep{Sackett1997}. This result is in agreement with the microlensing analysis of \citet{Wegg2016}.

\begin{table}
  \caption{Mass parameters of the best fitting model with uncertainties derived from the variation models of \autoref{section:stellarAndDarkMatterMass}.}
  \centering
  \begin{tabular*}{\columnwidth}{l  @{\extracolsep{\fill}} l}
    \hline
    \multicolumn{2}{l}{Mass inside a box of $(\pm 2.2 \times \pm 1.4 \times \pm 1.2)\kpc$ along the bar axes}\\
    \hline
    \Tstrut Smooth bulge traced by RCGs & $1.32 \pm 0.08 \times 10^{10}\, \Msun$\\
    Nuclear disk & $0.20 \times 10^{10}\, \Msun$\\
    Dark matter & $0.32 \pm 0.05 \times 10^{10} \, \Msun$\\
    \Bstrut Dynamical mass & $1.85\pm 0.05\times 10^{10}\Msun$\\
    \hline
    \multicolumn{2}{l}{Mass of the bulge, bar and inner disk}\\
    \hline	
    \Tstrut `Photometric'' bulge+bar & $1.88 \pm 0.12 \times 10^{10}\, \Msun$\\
    Inner disk & $1.29\pm0.12\times10^{10}\, \Msun$\\
    `Photometric' bulge & $1.34 \pm 0.04 \times 10^{10}\, \Msun$\\
    `Photometric' long bar & $0.54 \pm 0.04 \times 10^{10}\, \Msun$\\
    Non-axisymmetric long bar & $0.46 \pm 0.03 \times 10^{10}\, \Msun$\\
    \Bstrut Bar supporting orbits & $1.04 \pm 0.06 \times 10^{10} \, \Msun$\\
    \hline
    \multicolumn{2}{l}{Dark matter distribution}\\
    \hline
    \Tstrut Bulge dark matter fraction & $17 \pm 2 \%$\\
    Local dark matter density & $0.0132\pm0.0014\, \Msun.\pc^{-3}$\\
    Disk maximality at 2.2 disk scalelength & $75\%$\\
  \end{tabular*}
    \label{table:finalMassParameters}
\end{table}

\section{Discussion}
\label{section:discussion}

\subsection{An intermediate bar pattern speed}
\label{section:dicussionPaperI}

The bar pattern speed is a fundamental quantity that sets the position of resonances in the Milky Way. It has been the focus of many studies in the last two decades but the scatter between different determination methods remains surprisingly large. The new value determined in this paper from the dynamics of the bulge and long bar is $39\pm3.5\kmskpc$, based on several large photometric and kinematic survey data sets.

A number of previous determinations concluded on rather large pattern speeds in the range of $50-65\kmskpc$ from gas dynamics \citep{Englmaier1999, Fux1999, Bissantz2003}, continuity of a tracer stellar population \citep{Debattista2002} or interpretation of stellar stream in the solar neighborhood as a resonance effect \citep{Dehnen2000a,Antoja2014}. Other studies found lower values in the range $25-40\kmskpc$ from gas dynamics \citep{Weiner1999,Rodriguez-Fernandez2008, Sormani2015, Li2016} or stellar dynamics (\citealt{Long2013}; \hyperlink{P15}{P15}). 

Our new value for the bar pattern speed agrees very well with the extensive recent analysis of gas-dynamical models by \citet{Sormani2015}, but it is not consistent with the most precise determination so far based on recent analysis of the Hercules stream; this gives $53\pm0.5\kms$ when rescaled to $(R_0, V_0)=(8.2\kpc, 238\kms)$ \citep{Antoja2014}. However, this measurement is model dependent, assuming that the Hercules stream originates from the bar's outer Lindblad resonance. Future work must show whether an alternative interpretation of the Hercules stream can be found. 

Lower or intermediate pattern speeds are supported by the recent measurement of the long bar half-length of $5.0\pm0.2\kpc$ by \hyperlink{W15}{W15}. Since the bar cannot extend beyond corotation \citep{Contopoulos1980}, the corotation radius has to be greater than $\sim 5\kpc$, putting an upper limit on the pattern speed of about $\sim 48\kmskpc$. Bar pattern speeds are often expressed in terms of the dimensionless ratio $\mathcal{R}$ between the corotation radius $R_{\rm{cr}}$ and the bar half-length $R_{\rm{bar}}$. Typical values for external disk galaxies are $\mathcal{R} = 1.2\pm 0.2$ \citep{Elmegreen1996} with an indication of a correlation with bar strength \citep{Aguerri1998} and morphological type \citep{Rautiainen2008}. Using the measured $R_{\rm{bar}} = 5.0 \pm 0.2\kpc$ from \hyperlink{W15}{W15} and our determination of $R_{\rm{cr}} = 6.1\pm0.5\kpc$, we find for the Milky Way a ratio of $\mathcal{R} = 1.22\pm0.11$. This places the Milky Way together with the bulk of external barred spiral galaxies, with a bar that can be classified as a fast rotator ($\mathcal{R}\leq1.4$, \citet{Debattista2000}).

\subsection{The extra central mass}
\label{section:discussionCentralMass}

We found in \autoref{section:missingCentralMass} kinematic evidence for an additional central concentration of mass that was not included in our fiducial RCG bulge model. We can find good agreement with the kinematics by including an additional central mass component, mostly located in the plane where the RCG density has not been directly measured. Therefore, assuming that the required mass is stellar mass does not violate the 3D density measured from RCGs. In fact, there are several pieces of evidence for a stellar overdensity in the plane and near the GC but no consensus yet on the precise shape and mass of such overdensity. \citet{Launhardt2002} found a very concentrated nuclear bulge component in the central $220\pc$ from decomposition of the IRAS and COBE DIRBE data and estimate its mass to be $0.14\times 10^{10} \, \Msun$. From star counts, this component has a near-exponential vertical profile with scaleheight $45\pc$ corresponding to a nuclear stellar disk with axial ratio $\sim 3-5:1$ \citep{Nishiyama2013}. \citet{Schonrich2015} found in the \apogee data kinematic evidence of a nuclear disk extending to $\sim150\pc$ with a vertical height of $50\pc$. They measure a rotation velocity of $120\kms$ at $150\pc$ which, assuming an exponential density profile, leads to a nuclear bulge mass in reasonable agreement with the mass estimate from \citet{Launhardt2002}. In addition, \citet{Debattista2015} postulated the presence of a $\kpc$ scale nuclear disk to explain the high-velocity peaks in the line of sight velocity distributions of the \apogee commissioning data \citep{Nidever2012}. Such a large-scale disk would not be concentrated enough to sufficiently increase the central velocity dispersion required by the modelling, but could account for part of the mass in the galactic plane in the bulge region.

Hence, the most likely interpretation for our extra central mass component is simply a stellar over density near the centre, not included in the large-scale RCG bulge. This could well be a similar component as in the SBb galaxy NGC 4565 where HST and Spitzer photometry revealed an additional disky pseudo-bulge hidden inside the B/P bulge \citep{Kormendy2010}. More data closer to the centre and extending into the plane such as \apogee \citep{Majewski2012} or \gibs \citep{Zoccali2014} would be required to investigate the structure of this extra mass further. 

Some contribution to the required central mass could also come from metal poor stars not traced by RCGs such as old RR Lyrae stars \citep{Dekany2013, Pietrukowicz2015}. In the \argos fields, metal-poor stars with $\rm{[Fe/H]}\leq-0.9$ are only $7\%$ of the total sample but since they are more concentrated than the RCG bulge stars they could still play a role in the centre. However, there is still no clear picture about what component these stars belong to and what mass is associated with it. Favorite hypotheses are either an old thick disk \citep{DiMatteo2015}, a small classical bulge \citep{Kunder2016} or the inner part of the stellar halo \citep{Perez-Villegas2016}.

\subsection{Comparison with the bulge models of P15}
In \hyperlink{P15}{P15} we made a series of five dynamical models of the galactic bulge with different dark matter fractions called M80 to M90 by combining the 3D density of RCGs in the bulge from \citet{Wegg2013} and the \brava kinematics. Our main findings from these models were a measurement of the bulge total mass (stellar  + dark matter) of $1.84\pm0.07 \times 10^{10}\, \Msun$ and a rather low bulge pattern speed in the range $25-30\kmskpc$. In comparison, our new best model presented here has a total bulge mass of $1.85 \times 10^{10}\, \Msun$, in very good agreement with our previous determination. However, from our more advanced modelling, we find  a higher pattern speed of $40\kmskpc$. We attribute the difference to the lack of mass surrounding the bulge in the models of \hyperlink{P15}{P15}. As already stated in \autoref{section:InitialConditions}, bars formed in N-body models are generally several disk scalelengths long, which consequently underestimate the impact of inner disk and long bar orbits on the bulge dynamics. Hence, caution must be taken when interpreting the dynamics of N-body B/P bulges, unless the bulge, bar and disk scalelengths are all consistent with each other.

The present best model has a smooth bulge stellar mass of $1.32\times 10^{10}\, \Msun$, an additional central mass of $0.2\times 10^{10}\, \Msun$ and a dark matter mass in the bulge of $0.32 \times 10^{10} \, \Msun$, summing up to the $1.85\times 10^{10}\, \Msun$ stated above. Its closest equivalent in the \hyperlink{P15}{P15} series of models is the fitted model M82.5 which has a mass-to-clump ratio of $1014$, close to the value of 1000 we measured directly\footnote{This corrects Fig. 16 of \hyperlink{P15}{P15} where all model values of the mass-to-clump ratios are underestimated by $\sim 20\%$ due to an overestimate of the red giant branch bump fraction.}. The main differences are the larger bar pattern speed in the new model and the fact that the new model has $0.2\times 10^{10}\, \Msun$ less dark matter in the bulge but instead a similar additional mass close to the centre, as required by the central velocity dispersion for a larger pattern speed.

\subsection{Dark matter in the Galaxy}
\label{section:dicussionDM}

Early N-body dark matter simulations predicted a universal cuspy dark matter density profile \citep{Navarro1997}. This was later confirmed by new simulations with higher resolution able to resolve halo densities on scales smaller than a percent of $r_{200}$ \citep{Navarro2010}, i.e about $2\kpc$ for the Milky Way halo \citep{BlandHawthorn2016}. In addition, since the inner part of dark matter halos is actually populated by baryons that collapse through dissipative processes, dark halos should contract, thus exacerbating any pre-existing cusp \citep{Blumenthal1986, Gnedin2004, Abadi2009}. However, the degree of contraction of the halo varies greatly between authors (see \citealt{Wegg2016} for a recent comparison).

In disk galaxies, processes related to the dominant role of baryons in the central parts have been found to be able to transform a primordial cusp into a core. Such processes are the resonance effects of a large primordial stellar bar (\citealt{Weinberg2002}; but see \citealt{Dubinski2008}), supernova feedback \citep{Pontzen2012} and stellar feedback (\citealt{Chan2015}; \citealt{Schaller2014}; but see \citealt{Marinacci2013}). How important those processes are is not yet settled, resulting in a cusp/core controversy similar to that in dwarf galaxies \citep{Moore1994, Burkert1995, DeBlok2009}.

From our modelling, we find a best model with a low dark matter fraction in the galactic bulge of $17\pm2\%$ where the $2\%$ comes from the  variation models of \autoref{section:stellarAndDarkMatterMass}. In order to match simultaneously a low dark matter fraction in the bulge with the flat rotation curve close to the solar radius $R_0$, the dark matter density of our model has a power-law slope that is steeper than $\propto r^{-1}$ immediately inside $R_0$ and then flattens to a shallow cusp or a core in the bulge region. This result is consistent with the recent work of \citet{Wegg2016} who find that a high baryonic fraction is required to account for the high optical depths towards the bulge measured in the EROS-II and MOA-II microlensing data. Given the similarity between the best model here and the fiducial model of \citet{Wegg2016}, we expect our best model to be also consistent with microlensing constraints towards the bulge; this will be discussed in a later paper.

\section{Conclusion}
\label{section:conclusion}

We build a large number of dynamical models of the bar region in the Milky Way using the Made-to-Measure method. We first create a set of N-body models of barred disks that broadly matches the bulge, bar and outer disk density by adiabatic adaptation of a initial N-body model. This adiabatic procedure allows us to adapt the dark matter distribution to the rotation curve of the Galaxy inside 10\kpc and also to modify the pattern speed of the galactic bar. 
We then constrain those models with the stellar density of the bulge and bar as traced by red clump giants from a combination of the \vvv, \ukidss and \twomass surveys, together with stellar kinematics from the \brava, \ogle and \argos surveys. We explore a three-dimensional parameter space given by the stellar mass fraction in the bulge, the bulge and bar pattern speed and the nuclear disk mass, and provide constraints on the galactic effective potential. Our main conclusions are the following:
\begin{enumerate}
 \item Modelling the stellar dynamics in the bar region requires a bar pattern speed of $\Omega_b = 39\pm3.5\kmskpc$, placing corotation at $6.1\pm0.5\kpc$ from the Galactic Centre. The ratio of corotation radius to bar half-length ($R_{\rm{bar}} = 5.0 \pm 0.2\kpc$ from \citealt{Wegg2015}) of the Galaxy is found to be $\mathcal{R} = 1.22\pm0.11$, in good agreement with what is seen in external disk galaxies.
 
 \item We find a total dynamical mass of the bulge of $1.85\pm 0.05\times 10^{10}\Msun$, in excellent agreement with the value found in \citet{Portail2015a} from modelling the bulge only. 
 
 \item We find dynamical evidence for an extra central mass component, not included in our previous bulge models, of about $0.2 \times 10^{10} \, \Msun$ and probably related to a nuclear disk or disky pseudo-bulge.
 
 \item We evaluate from our model the mass of the long bar and bulge structure and find $M_{\rm{bar/bulge}} = 1.88 \pm 0.12 \times 10^{10}\, \Msun$, larger than the mass of disk in the bar region, $M_{\rm{inner\ disk}} = 1.29\pm0.12\times10^{10}\, \Msun$. The mass of the long bar is slightly less than half the disk mass in the same radial range. Our models predict a non-exponential surface density for the disk in the bar region and illustrate the transition between the bar region and the outer disk.
 
 \item We also evaluate the need of dark matter in the inner Milky Way. Using recent measurements of the bulge IMF and more extended data, we now better constrain the stellar-to-dark matter fraction in the bulge and find a preference for a mass-to-clump ratio of 1000 and a low dark matter fraction of $17 \pm 2\%$ in the bulge. In order to match simultaneously a low dark matter fraction in the bulge with the flat rotation curve close to the solar radius $R_0$, the dark matter density of our model has a power-law slope that is steeper than $\propto r^{-1}$ immediately inside $R_0$ and then flattens to a shallow cusp or a core in the bulge region.
\end{enumerate}
Our best-fitting model is the first non-parametric model of the entire bar region of the Milky Way and can be of significant use for several on going and future Milky Way studies including gas dynamics in realistic galactic bar potentials and chemodynamics of the different stellar populations. The model may be made available upon request to the authors.

\section*{Acknowledgements}
We thank Manuela Zoccali and Elena Valenti for helpful discussions, and are grateful to Jerry Sellwood and Monica Valluri for making their potential solver code available to us. We acknowledge the great work of the \vvv, \argos, \brava, \ukidss, \ogle and \twomass survey teams upon which this paper is built. We also thank the anonymous referee for their careful reading and constructive comments.


\phantomsection\label{lastpage}

\end{document}